\documentclass[12pt]{iopart}

\usepackage{geometry}
\usepackage{pdflscape}
\usepackage{bm}
\usepackage{natbib}
\usepackage{har2nat}   
\setcitestyle{aysep={,}} 
\usepackage{color}
\usepackage{hyperref}
\usepackage{mathrsfs}
\usepackage{graphicx}
\usepackage{epstopdf}
  \expandafter\let\csname equation*\endcsname\relax
  \expandafter\let\csname endequation*\endcsname\relax
\usepackage{amstext}
\usepackage{amsmath} 
\usepackage{eqnarray}
\usepackage{mathrsfs}
\usepackage{caption}
\usepackage[hang]{footmisc}
\usepackage{amssymb}
\usepackage{makecell}
\usepackage{xcolor}
\usepackage{tablefootnote}
\usepackage{multirow}
\usepackage{booktabs}
\usepackage{colortbl}
\hypersetup{
	colorlinks,
	linkcolor={red!50!black},
	citecolor={blue!50!black},
	urlcolor={blue!80!black}
}

\begin{document}

\title[Manuscript for Meas. Sci. and Technol.]{Error propagation dynamics of velocimetry-based pressure field calculations (2): on the error profile}

\author{Matthew Faiella$^{1}$, Corwin Grant Jeon Macmillan$^{1}$, Jared P. Whitehead$^{2}$ and Zhao Pan$^{1}$ }
\address{$^{1}$Department of Mechanical and Mechatronic Engineering, University of Waterloo, ON, Canada\\ 
$^{2}$Mathematics Department, Brigham Young University, UT, USA}
\ead{zhao.pan@uwaterloo.ca \& whitehead@mathematics.byu.edu}
\vspace{10pt}
\begin{indented}
\item[] March. 2021
\end{indented}

\begin{abstract}

A recent study investigated the propagation of error in a Velocimetry-based Pressure (V-Pressure) field reconstruction problem by directly analyzing the properties of the pressure Poisson equation \cite{Pan2016Error}.
In the present work, we extend these results by quantifying the effect of the error profile in the data field (shape/structure of the error in space) on the resultant error in the reconstructed pressure field. 
We first calculate the mode of the error in the data that maximizes error in the pressure field, which is the most dangerous error (called the worst error in the present work). 
This calculation of the worst error is equivalent to finding the principle mode of, for example, an Euler-Bernoulli beam problem in one-dimension and the Kirchhoff-Love plate in two-dimensions, thus connecting the V-Pressure problem from experimental fluid mechanics to buckling elastic bodies from elastic mechanics.
Taking advantage of this analogy, we then analyze how the error profile (e.g., spatial frequency of the error and the location of the most concentrated error) in the data field coupled with fundamental features of the flow domain (i.e., size, shape, and dimension of the domain, and the configuration of boundary conditions) significantly affects the error propagation from data to the reconstructed pressure. Our analytical results lend to practical applications in two ways. First, minimization of error propagation can be achieved by avoiding low frequency error profiles in data similar to the worst case scenarios and error concentrated at sensitive locations. Second, small amounts of the error in the data, if the error profile is similar to the worst error case, can cause significant error in the reconstructed pressure field; such a synthetic error can be used to benchmark V-Pressure algorithms.

\end{abstract}

\vspace{2pc}

\submitto{\MST}

\section{Introduction}
Velocity is the most pervasive measurement fluid-experimentalists use to gather information about flow fields. Various techniques over the past 20 years have supplied the fluids community with improved spatially and temporally resolved experimental data including: hot-wire anemometry, Laser Doppler Velocimetry (LDV), Particle Image Velocimetry (PIV), and Particle Tracking Velocimetry (PTV)) \cite{adrian2005twenty, Westerweel2013,dabiri2020particle}. Today's state-of-the-art systems can provide high-resolution volumetric velocity field data that can even compete with modern numerical methods \cite{Moin1998}. Velocity measurements are also fundamentally important in a wide range of industrial, military, medical and natural flow problems such as those associated with aircraft wings, shockwave interactions and vortex formation in prosthetic heart valves.

Although most techniques for non-invasive measurements have focused on the velocity field, extension to the pressure field has many promising aspects. Two review articles summarized the progress of the V-Pressure technique \cite{van2013piv,van2017comparative}. The improving accuracy and resolution of the velocimetry techniques promote velocimetry-based pressure field reconstruction techniques (V-Pressure) to be a more commonly used research tool to measure pressure fields in a non-invasive fashion. A few recent applications can be found in, for example, \cite{Zhang2019Combined,Zhang20194D,deem2020adaptive,pereira2020pressure}. 

Uncertainty quantification for the velocity field from techniques such as PIV has been significantly advanced \cite{wieneke2017piv,raffel2018particle,sciacchitano2019uncertainty}, but most modern experiments do not assess uncertainties in the pressure estimates and thus do not translate uncertainties in the velocity field measurements to the pressure field calculation. In an effort to clarify this issue, \citet{charonko2010assessment} benchmarked various PIV-based pressure calculation methods with numerical and physical experiments. They reported that the performance of the PIV-based pressure calculation is sensitive to many factors. Not surprisingly, they noted that several conditions (temporal and spatial resolution, velocity error, smoothing techniques, numerical scheme of the pressure solver, `flow type', etc.) impact the error propagation. In this pioneering research, they indicate that there is no universal or optimal method to reduce errors in the pressure calculation but that such reductions are case dependent. They did not report specifically on the effect of the error profile (e.g., spatial frequency of the error) in the velocimetry measurements on the resultant error in the calculated pressure field. 

\citet{de2012instantaneous} commented that the central finite difference based pressure Poisson solver acts as a low-pass filter, effectively eliminating the high-frequency errors in the pressure calculation. The ratio of the grid spacing of the numerical method to the temporal or spatial wave length of the experimental data impacts the frequency response of the pressure Poisson solver. Specifically, high frequency data (or noise) is filtered resulting in the loss of high-frequency physics (or low-pass filtering of the noise). Similarly, low frequency errors are more likely to propagate through the pressure calculation. 
% For example, a high Reynolds number turbulent flow field (high-frequency physics) with a calibration error in PIV (low frequency error) would result in large amount of error propagated to and dominated over the pressure field. 
\citet{de2012instantaneous} provides the first analysis on the error associated with frequency in the context of the V-Pressure; however, the results are limited to a specific numerical scheme. How the frequency response is affected by the `global' setup of the problem (e.g., configuration of boundary conditions (BC), size, and shape of the domain) was not discussed.

In a recent study, \citet{Pan2016Error} indicated that the profile of the error field in the data field ($\epsilon_f$) affects the error propagation.\footnote{`Data' is a commonly used term in applied mathematics, which represents the right hand side of a Poisson's equation and its BCs. To avoid confusion, we adopt this mathematical term in this paper hereafter as we did in \cite{Pan2016Error}, and address the velocity field measured from velocimetry experiments `velocimetry results' or `experimental results' instead.} 
For example, error in the data raised from peak-locking, calibration error, and random error, etc. may behave differently when propagating through a pressure solver. Their results analytically unravel the coupled effects of the error and fundamental features of the flow domain on the pressure calculation. They quantify how the error levels in the calculated pressure field  ($||\epsilon_p||_{L^2(\Omega)}$) can be bounded by the error in the data ($||\epsilon_f||_{L^2(\Omega)}$ and  $||\epsilon_h||_{L^\infty(\partial\Omega)}$ for error inside the domain and on boundary, respectively) and some Poincare constant(s) ($C_D$) which relate to the dimension and geometry of the domain (e.g., $||\epsilon_p||_{L^2(\Omega)}\leq  C_D||\epsilon_f ||_{L^2(\Omega)} + ||\epsilon_h||_{L^\infty(\partial\Omega)},$ for a domain with Dirichlet boundaries). However, this work does not identify a worst case error profile, or establish an error profile that will saturate the upper bound on the error in the calculated pressure field. In addition, the contribution of the profile of the error in the data field was not discussed in detail. A relevant work by \citet{gomit2018uncertainty} showed that when the velocimetry results are contaminated by different synthetic noise (Gaussian noise and modelled pixel-locking effect for example), the error propagation dynamics of V-Pressure  highly depends on the ``model of the errors''. In the context of the present research, the error profile (shape or structure in space) in the data is closely related to the ``model of the errors''. As implied in \cite{Pan2016Error} and explicitly suggested in \cite{gomit2018uncertainty}, error profile is important for uncertainty and error analysis for V-Pressure problems.

Here, we present a systematic methodology for finding the worst case error profile in the data (called the `worst error' hereafter) for a V-Pressure problem.
The procedure for finding the worst error directly exposes how the profile of the error in the data (e.g., spatial frequency and location of the error) coupled with the fundamental features of the flow domain (i.e., size, dimension, BCs) affects the error propagation. In addition, the calculation of the the worst error for the V-Pressure problem provides a surprising connection between fluid mechanics and elastic mechanics: the worst error profile for a V-Pressure problem is equivalent to the principle mode of a buckling Euler-Bernoulli beam in one-dimension (1D) and the Kirchhoff-Love plate in two-dimensions (2D). This analogy provides an analytical explanation how the profile of the error in the data affects the error in the reconstructed pressure field, in the same way as the bending moment applied on an elastic beam or plate affects the deflection of the elastic body. From a practical perspective, these results can be used to i) minimize error propagation by avoiding the worst case scenarios when designing and conducting V-Pressure experiments and ii) design the worst case benchmarking challenges to assess the performance of V-Pressure solvers.

The outline of the paper is as follows: first we \emph{define} the worst error possible for the V-Pressure calculation problem in \S\ref{sec:ProblemStatement}, then we \emph{calculate} the worst error in \S\ref{sec:CalculationOfTheWorstError}. In \S\ref{sec:Analogy} we elaborate the analogy between the (worst) error profiles for the  V-Pressure problem and the (principle) modes for a buckling elastic body using the Euler-Bernoulli beam as an example. The details of this analogy are summarized in table~\ref{tab:analogy}. Using this analogy, \S\ref{sec: impact of frequency of Ef} and \S\ref{sec: impact of  Ef location} demonstrate how the spatial frequency of the error and the location of a concentrated error source in the data field affect the error propagation of V-Pressure. We then provide analysis and examples in 1D and 2D in \S\ref{sec:1D example} and \S\ref{sec:2D examples} to show how the profile of the error in the data field affects the error propagation in a V-Pressure problem. We finally conclude the work in \S\ref{sec:Conclusions}.

\section{Problem statement}
\label{sec:ProblemStatement}
The propagation of error in the velocimetry-based pressure field calculation can be modeled via Poisson's equation 
\begin{equation}
\label{eq:ErrorPropagation}
\epsilon_f = \nabla^2 \epsilon_p, 
\end{equation}
where $\epsilon_f$ and $\epsilon_p$ are the error in the data field and calculated pressure field, respectively \cite{Pan2016Error}. The error level is measured by the $L^2$ norm; for example, the error level of the calculated pressure field is 
\begin{equation}
\label{eq:L2Norm}
||\epsilon_p||_{L^2(\Omega)} = \sqrt{\frac{\int \epsilon_p^2 d\Omega}{|\Omega|}}, 
\end{equation}
where $\Omega$ is the domain of the flow field, and $|\Omega|$ is the length, area or volume of the domain, depending on the dimension. 
This $L^2$ norm is intuitively a measurement of the space-averaged power of the error. 
This norm is similar to the root mean square (RMS) in a discrete context, although we emphasize here that the current discussion is independent of the choice of numerical discretization. 
The major goal of the current research is to determine the error profile in the data field ($\epsilon_f$) that leads to the worst error (measured by this norm) in the calculated pressure field ($\epsilon_p$) relative to the error in the data itself. 
Thus, finding the worst error can be defined as a variational problem with $\epsilon_f$ as the desired function, i.e., we seek $\epsilon_f$ (with a particular profile) to satisfy:
\begin{equation}
\label{eq:OptimizationProblem}
Ar^*=\max_{\epsilon_f} Ar =  \max_{\epsilon_f} \frac{||\epsilon_p||_{L^2(\Omega)}}{||\epsilon_f||_{L^2(\Omega)}},
\end{equation}
subject to \eqref{eq:ErrorPropagation} with the appropriate BCs on the domain.  $Ar$ is the ratio of the error level in the reconstructed pressure field relative to the error level in the data (i.e. $Ar =   \frac{||\epsilon_p||_{L^2(\Omega)}}{||\epsilon_f||_{L^2(\Omega)}}$). 
The interpretation of this optimization problem is that we seek an unfriendly error in the data field ($\epsilon_f$) taking a particular shape or \textit{profile}, with a constant \textit{power} ($||\epsilon_f||_{L^2(\Omega)}$), that negatively affects the pressure solver the most, simply due to its special profile, so that the error amplification ratio is maximized. 
If we consider the data field as the input of a V-Pressure solver and the pressure field as the output, $Ar$ is an error amplification ratio, and $Ar^*$ is the highest possible value of this error amplification factor.
Hereafter, we address $Ar$ as the error amplification ratio, and address the solution of the optimal problem \eqref{eq:OptimizationProblem} as the \textit{worst error}, denoted as $\epsilon_f^*$. 
We also consider the worst error to correspond to the most dangerous mode of $\epsilon_f$, which is amplified through solving a pressure Poisson equation. 
We are only interested in the situations where the denominator in~\eqref{eq:OptimizationProblem} is non-zero as this indicates that the power in the error in the data is nonzero, thus avoiding the mathematical singularity in $Ar$ and reflecting the reality that all experimental data will have error.

\section{Calculation of the worst error for pressure reconstruction}
\label{sec:CalculationOfTheWorstError}
Here we first consider the maximization problem of \eqref{eq:OptimizationProblem} with Dirichlet BCs on the data $\epsilon_f$ (Neumann and mixed BCs can be treated similarly).  Using the error Poisson equation \eqref{eq:ErrorPropagation} we can simplify $Ar^*$ to be in terms of the output error only:
\begin{equation}
\label{eq:OptimizationProblem2}
Ar^*=\max_{\epsilon_p}  \frac{||\epsilon_p||_{L^2(\Omega)}}{||\nabla^2 \epsilon_p||_{L^2(\Omega)}}, \quad  \epsilon_p=g ~\textrm{for} ~x \in  \partial \Omega,
\end{equation}
where $g$ is a sufficiently smooth function that specifies the boundary condition. 
This can be reformulated using a constant Lagrange multiplier with the constraint that the denominator is normalized (see \citet{Gelfand1991} for example). 
In other words, we seek the solution of:
\begin{equation}
\begin{aligned}
\label{eq:lagrange1}
& \underset{\epsilon_p}{\text{max}}& & J[\epsilon_p] = \max_{\epsilon_p} \int_\Omega \left\{ |\epsilon_p|^2 + \lambda |\nabla^2 \epsilon_p|^2\right\}dx,\\ 
& \text{s.t.}& & \int_\Omega |\nabla^2 \epsilon_p|^2 dx = 1,\quad\mbox{and}\quad \epsilon_p=g~\textrm{for}~x\in\partial\Omega.
\end{aligned}
\end{equation}
Standard application of the calculus of variations (see \citet{Gelfand1991}) then indicates that the maximizer for this problem must satisfy the Euler-Lagrange equations:
\begin{equation}
\label{eq:4thOrderGeneral}
\nabla^4 \epsilon_p = -\frac{1}{\lambda}\epsilon_p,
\end{equation}
with $\epsilon_p=g$ and $\nabla^2 \epsilon_p = 0$ on $\partial\Omega$, subject to $|\Omega|\|\nabla^2 \epsilon_p\|_{L^2(\Omega)}^2 = 1$.

The additional boundary condition $\nabla^2\epsilon_p = 0$ is the `natural' boundary condition that arises because the variational formulation is quadratic in the second derivatives of $\epsilon_p$, and we have specified only the first order Dirichlet condition from physical considerations.  As indicated above, other boundary conditions (other than Dirichlet BCs) will result in a different type of natural boundary conditions as listed in table~\ref{tab:BCs} (see also \citet{Gelfand1991}).

\begin{table}
	\centering
\caption{\label{tab:BCs} Type of boundary conditions (BCs) of the original pressure Poisson equation, the corresponding BCs of the eigenvalue problem of the worst error and the induced natural boundaries. $G$, $g$, $H$, and $h$ are functions on the boundary $\partial \Omega$, and $\hat{n}$ is the unit outward pointing normal on $\partial\Omega$.} 
	\footnotesize
	\begin{tabular}{llll}
		\br
		{Type of BCs}  & {\makecell{BCs of pressure \\ Poisson equation}} & {\makecell{Essential BCs of\\ eigenvalue problem}}  & {\makecell{Natural BCs of\\ eigenvalue problem}}\\
		\mr
		{Dirichlet}  &	$p = G$ &	$\epsilon_p = g$     & $  \nabla^2 \epsilon_p =0 $ \\
		{Neumann} &	$\nabla p\cdot \hat{n} = H$ &	$\nabla\epsilon_p\cdot \hat{n} = h$     & $  \nabla\left(\nabla^2 \epsilon_p\right) \cdot\hat{n} =0 $ \\
		\br
	\end{tabular}
\end{table}
\normalsize

Equation~\eqref{eq:4thOrderGeneral} is the same as the eigenvalue problem that arises as the characteristic equation of transverse vibration of beams or plates \cite{timoshenko1937vibration}.
Thus, as long as the boundary conditions are prescribed carefully, i.e. the operator remains self-adjoint (as they are for the Dirichlet case specified here), we are guaranteed that there is a countable number of solutions to this system with corresponding natural frequencies $\beta_n = \sqrt[4]{-\lambda_n^{-1}}$, where $\lambda_n>0,  n = 1,2,3...$ are the eigenvalues. 
The smallest of these eigenvalues $-\lambda_1^{-1}$ and corresponding eigenfunction yield the desired profile $\epsilon^*_f$, which is the worst error in the data and leads to the highest possible error amplification ratio $Ar^*$.
In other words, the worst error in the data field is the eigenfunction (or the principle mode) that is associated with the first eigenvalue, and this worst error in the data field will be amplified most and leads to highest possible $Ar$, which is $Ar^*$.
Fundamental features of the flow domain (i.e., dimension, size and shape of the domain, and the configuration of BCs) determine the eigenvalue (see also \citet{Pan2016Error}). Thus, the worst error profile in the data ($\epsilon^*_f$) is indeed determined by the fundamental features of the flow domain. 
Similarly, the second worst error is associated with the second eigenvalue ($-\lambda_2^{-1}$), which is larger than the first eigenvalue ($-\lambda_2^{-1}>-\lambda_1^{-1}$) and leads to the second largest $Ar$.
The corresponding eigenfunction has higher spatial frequency, and yields a value of $Ar$ that is lower than the worst error, and so on for the third worst error. This analysis lends insight into how the error propagation is affected by spatial frequency of error in the data, which will be discussed in later sections.

This fourth order eigenvalue problem resulting from the fourth order variational problem may not be trivial to the fluid mechanics community, and the calculations can be tediously long even for one dimensional (1D) cases. However, it has been studied in great detail in solid mechanics (e.g., \citet{timoshenko1937vibration,landau1986course}). Specifically, it is equivalent to the Euler-Bernoulli beam problem in 1D, and the Kirchhoff-Love plates problem in 2D. Solutions with standard boundary conditions and in basic domains are tabulated in popular textbooks (e.g. \citet{morse1948vibration}, \citet{harris2002harris}), and convenient to look up for fluid mechanics researchers with help from table~\ref{tab:BCs}. In the following section, we provide examples to demonstrate the analogy.

\section{Analogy: from buckling elastic bodies to V-Pressure error propagation}
\label{sec:Analogy}
In this section, we elaborate the analogy between the buckling of elastic bodies and the error propagation dynamics of V-Pressure calculation using the Euler-Bernoulli beam as an example in 1D. We first provide a brief review of the classic theory of the Euler-Bernoulli beam and the resulting eigenvalue problem. Next, a Poiseuille flow is used to demonstrate how error in the data field ($\epsilon_f$) is propagated to the calculated pressure ($\epsilon_p$) through a pressure Poisson solver, and the same eigenvalue problem is solved to find the worst error. Both heuristic interpretation and rigorous discussion are provided to illustrate the connections between the beam problem and the velocimetry-pressure problem. In table~\ref{tab:analogy}, we 
summarize the analogy side by side by listing the physical meanings of the relevant equations and terms, in the context of elastic mechanics and fluid mechanics respectively, and how they are connected by shared mathematical roots.

\subsection{Vibration and buckling of an Euler-Bernoulli beam}
\label{sec:Analogy - elastic body}
Consider a homogeneous beam of length $L$ which undergoes transverse vibrations. The normalized deflection profile of the beam is governed by the differential equation:
\begin{equation}
    \label{eq:E-B Beam}
     \frac{d^4Y}{dx^4} = \frac{d^2Y}{dt^2}, 
\end{equation}
where $Y(x,t)$ is the beam deflection in the direction that is perpendicular to the $x$~coordinate, and $t$ is time. By separation of variables ($Y(x,t)=X(x)T(t)$), where $T(t)$ is a function of $t$ and $X(x)$ is a function of the spatial $x$ coordinate and hence describes the modes of the vibrating beam, the corresponding eigenvalue problem of \eqref{eq:E-B Beam} is:
\begin{equation}
    \label{eq:E-B BeamEigenfun}
     \frac{d^4X}{dx^4} = \gamma X,  
\end{equation}
where $\gamma$ is the eigenvalue, and $k=\sqrt[4]{\gamma}$  \cite{timoshenko1937vibration} is the spatial frequencies of the buckled beam. Different eigenvalues ($\gamma_n=\omega_n^{-2}, n = 1,2,3,...$) correspond to different natural frequencies ($\omega_n$) of the vibrating beam and the eigenfunctions ($X_n$) describing the modes of the beam deflections. Larger eigenvalues correspond to beam modes with higher spatial frequencies.
We also recall the normalized governing equations of a bending beam:
\begin{align}
\label{eq:E-B Beam eqs 1}
     \frac{dX}{dx} & =  \Theta,  \\ 
\label{eq:E-B Beam eqs 2}
     \frac{d^2X}{dx^2} & = - M, \\ 
\label{eq:E-B Beam eqs 3}
     \frac{d^3X}{dx^3} & = - Q, \\
\label{eq:E-B Beam eqs 4}
     \frac{d^4X}{dx^4} & = - W,
\end{align}
where $\Theta$ is the slope, $M$ is the bending moment, $Q$ is the shear force in the beam, and $W$ is the transverse load on the beam, respectively. 

One may notice that \eqref{eq:E-B BeamEigenfun}, which is derived from the beam problem, is identical to the one-dimensional version of \eqref{eq:4thOrderGeneral}, which arises in the error propagation problem of the pressure field calculation based on velocimetry data. In addition, \eqref{eq:E-B Beam eqs 2} describes the relationship between the second order derivative of the beam deflection and bending moment; and it  takes the form of the Poisson's equation about error propagation, i.e., \eqref{eq:ErrorPropagation}, in one dimension. This indicates an analogy between the source terms of \eqref{eq:E-B Beam eqs 2} and \eqref{eq:ErrorPropagation}: the bending moment ($M$) exerted on a beam causing the buckling of the beam for the elastic mechanics problem is equivalent to the error in the data ($\epsilon_f$) contaminating the pressure Poisson equation for the V-Pressure error propagation problem. Similarly, the deflection profile of the beam ($X$) is equivalent to the error in the calculated pressure field ($\epsilon_p$).

\begin{figure}
	\centering 
	\includegraphics[width=0.5\textwidth]{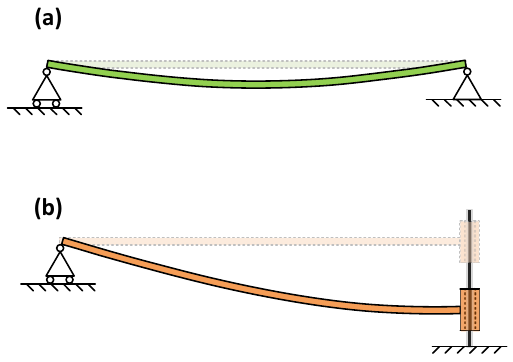}
	\caption{Bent Euler-Bernoulli beam with different supporting mechanisms: (a) simply supported at both ends (associated with Dirichlet-Dirichlet BCs) and (b) simply supported at the left end and supported by a slide at the right end (associated with Dirichlet-Neumann BCs).}
	\label{fig:beam}
\end{figure}

Solving~\eqref{eq:E-B BeamEigenfun} requires four boundary conditions. For example, if both ends of the beam are simply supported (also called hinged in some textbooks, and see Figure~\ref{fig:beam}(a) for illustration), the displacement at the ends are constrained by the kinematic boundary conditions (also called essential boundary conditions): 
\begin{equation}
\label{eq: essential BC hinged}
    X(0)=X(L)=0, 
\end{equation}
and natural boundary conditions indicating that no bending moments are exerted at the ends (see \citep{harris2002harris} and \eqref{eq:E-B Beam eqs 2}): \begin{equation}
\label{eq: natrual BC hinged}
    X''(0)=X''(L)=0.
\end{equation}
Equipped with \eqref{eq: essential BC hinged} and \eqref{eq: natrual BC hinged}, the eigenvalue problem \eqref{eq:E-B BeamEigenfun} of the Euler-Bernoulli beam is ready to be solved to find the deflection of the beam. The solution procedure is trivial; however, it will be showed in the following section (\S\ref{sec:Analogy - PIVPressure}) in the context of error propagation of V-Pressure to further demonstrate the analogy.

\subsection{Error propagation of pressure reconstruction of a Poiseuille flow}
\label{sec:Analogy - PIVPressure}
In the context of error propagation of V-Pressure, we now compute the worst error in the data field ($\epsilon^*_f$) and determine how it propagates to the calculated pressure field to generate the highest possible error amplification ratio ($Ar^*$).  To do this, we solve the eigenvalue problem \eqref{eq:4thOrderGeneral}. 
Consider the center-line of a Poiseuille flow on $x\in[0,L]$, the pressure field of the flow can be calculated by solving a 1D `Poisson's equation' ($p''=0$) with two Dirichlet BCs at $x=0$ and $x=L$ (see the example in \S\ref{sec:1D example} for more details).
The corresponding error function is $\epsilon_p''=\epsilon_f$, which is the 1D form of \eqref{eq:ErrorPropagation}. 
Assuming error in the data field ($\epsilon_f$) is not negligible and dominates over the error on the Dirichlet boundaries, the 1D case of Poiseuille flow reduces the eigenvalue problem of (\ref{eq:4thOrderGeneral}) to:
\begin{equation}
\label{eq: 1D4thOrderEigenvalue}
\frac{d^4 \epsilon_p}{dx^4} = -\frac{1}{\lambda}\epsilon_p, 
\end{equation}
and the essential boundary conditions of the eigenvalue problem come from the \textcolor{black}{homogeneous}\footnote{\color{black}Nonhomogeneous BCs arise from scenarios where the impact of the error on the boundaries cannot be ignored. For example, they occur when the size of the domain is small and/or the error on the boundary is not negligible. Exact analysis on a domain with nonhomogeneous BCs can be significantly more difficult. However, numerical investigation similar to the simulations demonstrated in \S~\ref{sec:Effects of the location of ef in 2D} can be used to analyze error sensitivity regardless of the error on the boundaries.} Dirichlet BCs of the pressure Poisson problem:
\begin{equation}
   \label{eq: essential BC DBC flow}
   \epsilon_p(0) = \epsilon_p(L) = 0. 
\end{equation}
The corresponding natural BCs from the Calculus of Variations (see also table~\ref{tab:BCs}) are 
\begin{equation}
    \label{eq: natural BC DBC flow}
   \epsilon''_p(0) = \epsilon''_p(L) = 0.
\end{equation}
At this juncture we emphasize the similarity between the eigenvalue problems \eqref{eq: 1D4thOrderEigenvalue} and \eqref{eq:E-B BeamEigenfun}, and the comparison between essential boundary conditions \eqref{eq: essential BC DBC flow} and \eqref{eq: natrual BC hinged}, as well as the natural boundary conditions \eqref{eq: natural BC DBC flow} and  \eqref{eq: essential BC hinged}. 

The general solution of \eqref{eq: 1D4thOrderEigenvalue} is $\epsilon_p(x) =C_1\cos (\beta x) + C_2\sin (\beta x)    + C_3\cosh (\beta x) + C_4 \sinh (\beta x),$ where $C_i~(i=1\dots4)$ are constants to be determined.
Applying the boundary conditions \eqref{eq: essential BC DBC flow} and \eqref{eq: natural BC DBC flow}, we find $\beta_n=n\pi/L, n = 1,2,3,...$ and the corresponding eigenfunctions, which are the modes of the error profile in the pressure field:
\begin{equation}
    \label{eq:modes in Ep}
    {\epsilon_{p}}_n=\sin(\beta_n x).
\end{equation}
The above solution also reflects a typical procedure for solving a beam problem such as \eqref{eq:E-B BeamEigenfun}. The modes of a simply supported beam take the identical form as in \eqref{eq:modes in Ep} due to exactly the same calculation. Recalling the Poisson equation \eqref{eq:ErrorPropagation}, differentiation \eqref{eq:modes in Ep} twice gives the modes of the error profile in the data:
\begin{equation}
    \label{eq:modes in Ef}
    \epsilon_{{f}_n}=\epsilon''_{{p}_n} =-\beta_n^2{\epsilon_p}_n.
\end{equation}

Normalizing the power of the error in the data field ($\|\epsilon_f(x)\|_{L^2(\Omega)} =1 $) the exact profile $\epsilon_f$ can be found for each mode:
\begin{equation}
    \label{eq: worst error Ef}
    {\epsilon_{f}}_n = \pm \sqrt{2}\sin\left(\frac{n\pi}{L}x\right),
\end{equation} 
which will yield corresponding error in the calculated pressure field:
\begin{equation}
    \label{eq: worst error Ep}
    {\epsilon_{p}}_n = \pm \sqrt{2}\frac{L^2}{n^2\pi^2}\sin\left(\frac{n\pi}{L}x\right).
\end{equation} 
The maximum of the error amplification ratio $Ar^*$ occurs for $n=1$ wherein the maximal error in the pressure field, $\epsilon_p =  \pm \sqrt{2}\frac{\pi^2}{L^2} \sin(\frac{\pi}{L} x)$, is induced by an error in the data field with the specific profile:
\begin{equation}
    \label{eq: worst error in data field DD}
    \epsilon^*_f = \pm \sqrt{2}\sin\left(\frac{\pi}{L} x\right), 
\end{equation}
and this $\epsilon^*_f$ is the worst error in the data field we are looking for. The highest possible error amplification ratio is
\begin{equation}
    \label{eq: Ar* - DD}
    Ar^*=\frac{L^2}{\pi^2},
\end{equation}
for a such pressure field reconstruction problem with Dirichlet BCs at both endpoints. When carrying out velocimetry experiments, one should avoid any error having a similar profile to $\epsilon^*_f$ (essentially high error in the middle of the domain and far from the tied down boundaries), which is particularly unfriendly to a pressure Poisson solver. A detailed demonstration will be provided in \S\ref{sec:1D example}.

\newgeometry{margin= 1in} 
\begin{landscape}
\begin{table}
\centering
\caption{Analogy between the beam vibration problem and the error propagation problem raised by the Velocimetry-based pressure reconstruction.}
\small
\arrayrulecolor[rgb]{0.753,0.753,0.753}
\label{tab:analogy}
\begin{tabular}{ccccc} 
\arrayrulecolor{black}\toprule
Physical interpretation                                                              & \begin{tabular}[c]{@{}c@{}}Euler–Bernoulli~beam \\vibration problem\end{tabular} & Mathematics                                                                                                                                & \begin{tabular}[c]{@{}c@{}}Velocimetry-pressure error \\propagation problem\end{tabular} & Physical interpretation                                                                                       \\ 
\midrule
\begin{tabular}[c]{@{}c@{}}Vibration motion of a\\ Euler-Bernoulli beam\end{tabular}      & $Y''''=-\ddot{Y}$                                                                         & Original governing eq.                                                & $\nabla^2p=f$                                                                            & Pressure Poisson eq.                                                                                          \\ 
\arrayrulecolor[rgb]{0.753,0.753,0.753}\hline
\begin{tabular}[c]{@{}c@{}}Beam deflection due\\~to bending moment \end{tabular} & $ X''=-M$                                                                 & Induced governing ~eq.\tablefootnote{The governing equations are normalized to expose the mathematical roots shared by the two problems. For the beam problem, $Y(x,t)=X(x)T(t)$ is the deflection of the beam, which is a function of $x\in[0,L]$, and time $t\in[0,\infty)$. $[~]^{'}$ indicates the derivative with respect to $x$, and $\ddot{[~]}$ indicates the time derivative. For the error propagation problem in V-Pressure, derivation can be found in \cite{Pan2016Error}.} & $\nabla^2\epsilon_p=\epsilon_f$                                                          & Error propagation eq.                                                                    \\ 
\hline
\begin{tabular}[c]{@{}c@{}}Eq. determining\\~modes of the beam\end{tabular}          & $X''''= \gamma X$                                                                     & \makecell{4th order\\ eigenvalue problem\tablefootnote{The derivations for the Eigenvalue problem can be found in \citet{timoshenko1937vibration} (for the beam problem) and in Sec.~\ref{sec:CalculationOfTheWorstError} (for the PIV-pressure error problem).}}                                          & $\nabla^4\epsilon_p=-\lambda^{-1}\epsilon_p$                                             & \begin{tabular}[c]{@{}c@{}}Euler-Lagrange eq. \\determining the worst\\~error in pressure field~\end{tabular}  \\ 
\hline
\begin{tabular}[c]{@{}c@{}}Eigenvalues for \\bending beam\end{tabular}               & $\gamma_n$                                                                            & Eigenvalues                                                            & $-\lambda_n^{-1}$                                                                  & \begin{tabular}[c]{@{}c@{}} Eigenvalues determined by\\the Lagrange multipliers\end{tabular}       \\ 
\hline
\begin{tabular}[c]{@{}c@{}}The 1st, 2nd,... modes \\of beam deflection\end{tabular}  & $X_{n},~n = 1,2,...$                                                             & Eigenfunctions                                                         & ${\epsilon_{p}}_n,~n = 1,2,...$                                                            & \begin{tabular}[c]{@{}c@{}}The worst, 2nd worst…\\error~in the pressure field\end{tabular}                    \\ 
\hline
\begin{tabular}[c]{@{}c@{}}The 1st, 2nd,...~modes~\\of bending moment\end{tabular}     & $X''_{n},~n = 1,2,...$                                                           & \begin{tabular}[c]{@{}c@{}} 2nd order derivative \\of eigenfunctions\end{tabular}          & $\nabla^2\epsilon_{p,n}={\epsilon_{f}}_n,~n = 1,2,...$                                     & \begin{tabular}[c]{@{}c@{}}The worst, 2nd worst…~ \\error in the data field\end{tabular}                      \\ 
\hline
\multirow{2}{*}{Hinged end of a beam~}                                               & $X(0)=0$                                                                         & Essential BC\tablefootnote{Homogeneous boundary conditions at $x=0$ are used as examples.}                                                                                                                                                                                                                                                                                                                                                                                                     & $\epsilon_p(0)=0$                                                                        & \multirow{2}{*}{\begin{tabular}[c]{@{}c@{}}Dirichlet BC enforced\\on~a flow domain \end{tabular}}             \\ 
\cline{2-4}
                                                                                     & $X''(0)=0$                                                                       & Natural BC                                                                   & $\epsilon_p''(0)=0$                            &                                                                                                 \\
                                                                                     \hline
\multirow{2}{*}{\begin{tabular}[c]{@{}c@{}}Slide-supported \\end of a beam \end{tabular}}                                               & $X'(0)=0$                                                                         & Essential BC                                                                                                                                                                                                                                                                                                                                                                                                   & $\epsilon_p'(0) = 0$                                                                        & \multirow{2}{*}{\begin{tabular}[c]{@{}c@{}}Neumann BC enforced\\on~a flow domain \end{tabular}}             \\ 
\cline{2-4}
                                                                                     & $X'''(0)=0$                                                                       & Natural BC                                                                   & $\epsilon_p'''(0)=0$                            &                                                                                                 \\
\arrayrulecolor{black}\bottomrule
\end{tabular}
\end{table}
\normalsize
\end{landscape}
\restoregeometry

% \subsection{Effects of BC configuration on a domain}
% \section{Combined effects of BC and spatial frequency of the error}
\section{Impact of the spatial frequency of $\epsilon_f$}
\label{sec: impact of frequency of Ef}
In this section, we elaborate on how the spatial frequency of the error in the data, coupled with the fundamental features of the domain, impacts the error propagation from $\epsilon_f$ to $\epsilon_p$. When Dirichlet BCs are enforced, \eqref{eq: Ar* - DD} gives the worst possible error amplification ratio $Ar^*$. 
% The calculation of \eqref{eq: Ar* - DD} illustrates how the spatial frequency of the error in the data couple with the fundamental features of the domain affects the error propagation. 
In fact, before arriving at \eqref{eq: Ar* - DD}, rearranging~\eqref{eq:modes in Ef} leads to
\begin{equation}
\label{eq: Ar(n) - DD}
    Ar=\beta_n^{-2} = \frac{L^2}{\pi^2 n^2}, \quad n = 1,2,3,...
\end{equation}
which means that the error amplification ratio ($Ar$) for each mode (or frequency component) of $\epsilon_f$ directly relates to the eigenvalues of the problem, and thus the fundamental features of the domain. \eqref{eq: Ar(n) - DD} indicates that low frequency error in a large domain (i.e., small $n$ and large $L$) are amplified than high frequency error.
% This results concurs the findings in \citet{Pan2016Error}.

The error amplification rate in \eqref{eq: Ar(n) - DD} is developed based on a flow field with Dirichlet-Dirichlet boundaries, which is analogous to a vibrating beam with two simply supported ends (see Figure~\ref{fig:beam}(a)) in the context of elastic mechanics. It is expected that such a `firm support' (compared to, for example, a beam with a free end) prevents excessive bending of the beam. Returning to the context of pressure field reconstruction, introducing accurate Dirichlet BCs for the domain (as `support') can effectively tame the error propagated to the pressure field from the data. One example of this in practice can be found in the recent work of \cite{shanmughan2020optimal}. 

% We now investigate a problem on a domain with Dirichlet-Neumann boundaries. 
For the same 1D pressure reconstruction problem for the aforementioned Poiseuille flow , if mixed BCs are enforced on the pressure Poisson equation (e.g., Dirichlet BC at $x=0$, and Neumann BC at $x=L$), the resulting BCs of the eigenvalue problem \eqref{eq: 1D4thOrderEigenvalue} is the same as \eqref{eq: essential BC DBC flow} and \eqref{eq: natural BC DBC flow} at $x=0$. The Neumann BC at $x=L$ leads to an essential BC of $\epsilon'_p(L) = 0$ and its natural BC $\epsilon'''_p(L) = 0$. Similar calculation to the Dirichlet-Dirichlet case yields
\begin{equation}
\label{eq: Ar(n) - DN}
Ar = \frac{4L^2}{\pi^2(2n-1)^2}, n = 1,2,3...
\end{equation}
and the first eigenvalue corresponds to the worst error in the data field $\epsilon_p = \pm \sin(\frac{\pi}{2L}x)$, which is achieved for $n=1$. Thus, the highest possible error amplification ratio for the problem with mixed BCs is
\begin{equation}
    \label{eq: Ar* - DN}
    Ar^* = 4\frac{L^2}{\pi^2}. 
\end{equation}

\begin{figure}[h!]
	\centering 
	\includegraphics[width= 0.5 \textwidth]{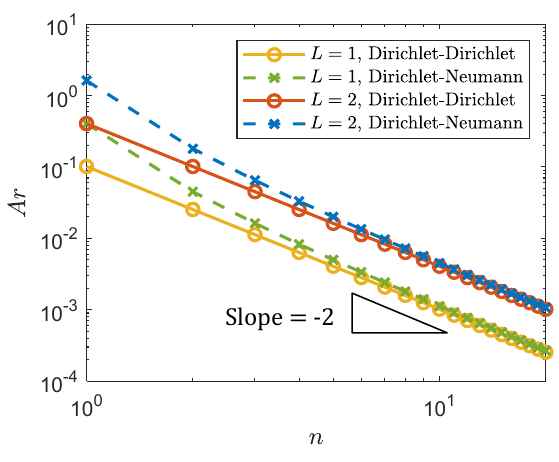}
	\caption{Amplification ratio ($Ar$) as a function of the spatial frequency of the error in the data, when the domain has different size ($L=1$ or $2$) with different BCs (Dirichlet-Dirichlet or Dirichlet-Neumann).}
	\label{fig:Ar(n)}
\end{figure}

Both \eqref{eq: Ar(n) - DD} and \eqref{eq: Ar(n) - DN} indicate that the spatial frequency of the error in the data field ($\epsilon_f$), coupled with the fundamental features of the domain, affects the error propagation from $\epsilon_f$ to $\epsilon_p$ and leads to a varied amplification ratio ($Ar$). Figure~\ref{fig:Ar(n)} illustrates $Ar$ as a function of $n$, for the domain with different sizes (e.g., $L=1$~or~$2$) and different boundary condition configurations (e.g., Dirichlet-Dirichlet for \eqref{eq: Ar(n) - DD} and Dirichlet-Neumann for \eqref{eq: Ar(n) - DN}). We note that the configuration of BCs affects the error propagation. For a given domain with fixed size, a Dirichlet-Neumann boundary setup amplifies $\epsilon_f$ more than the Dirichlet-Dirichlet case when the spatial frequency of $\epsilon_f$ is low. For example, as $n \rightarrow 1$, $Ar \rightarrow Ar^*$, and $Ar^*$ for the Dirichlet-Neumann case is four times larger than the Dirichlet-Dirichlet case. In the context of the Euler-Bernoulli beam theory, a Neumann BC of the error propagation problem leads to a `beam' supported by a slide, where only the slope, not the deflection, at the end of the beam is enforced to be zero, and introduces a natural BC only indicating a vanishing shear force at the end (see Figure~\ref{fig:beam}(b) for beam supported by a slide at right end). With such a  `weak support', one can expect more overall deflection across the entire beam, and excessive deflection at the slide support. Returning to the context of V-Pressure, it indicates that more error in the calculated pressure field (and high error near Neumann boundaries) is expected when Neumann BCs present.

In the limit as $n \rightarrow \infty$, \eqref{eq: Ar(n) - DN} approaches \eqref{eq: Ar(n) - DD} (see also Figure~\ref{fig:Ar(n)}, the dashed lines collapse on the solid lines for each domain size when $n$ is large). This means that the effects of BCs on high frequency errors vanish, and the size of the domain ($L$) and spatial frequency ($n$) together dominate the error propagation. 
For example,  \eqref{eq: Ar(n) - DD} indicates $Ar \sim L^2$, meaning that error in the data is amplified more by a flow domain with larger length scale $L$, which induces more error in the calculated pressure field ($\epsilon_p$). From \eqref{eq: Ar(n) - DD}, we also note $Ar \sim n^{-2}$, meaning that high spatial frequency components of the error in the data field (associated with large $n$) are less amplified through the pressure reconstruction.  To be more specific, $Ar$ decays at a power rate of $-20$ dB/decade, as if the high frequency modes are filtered by a low-pass filter with a such roll-off.
For the Euler-Bernoulli beam (see table~\ref{tab:analogy}), we note that a longer beam tends to bend more, and a higher frequency vibration typically associates with lower amplitude or deflection.

The low-pass filtering effect illustrated here is independent of the numerical scheme that is used to implement the pressure Poisson solver and the velocimetry technique. Here, we want to emphasize that the frequency response analysis in the current work is different than some results, for example, presented in \citep{de2012instantaneous}, where a `local' analysis was employed: the amplitude response of a signal or error passing through a Poisson solver is given by a transfer function
\begin{equation}
\label{eq:deKat12 trans fun}
 T_{PS}(h,\lambda_x) = \frac{|p|}{|f|} = \frac{|\epsilon_p|}{|\epsilon_f|} =  \frac{1+\cos\left( \pi\frac{2h}{\lambda_x}\right)}{2~\text{sinc}\left(\frac{2h}{\lambda_x}\right)}, 
\end{equation}
where $T_{PS}$ is the transfer function of the Poisson solver, $h$ is the grid spacing, and $\lambda_x$ is the spatial wavelength of the input signal (or error) to the numerical Poisson solver, respectively. In the limit of $h/\lambda_x\rightarrow 0$, according to \eqref{eq:deKat12 trans fun}, $ T_{PS}(h,\lambda_x)\rightarrow1$. This result considers numerical implementation of the pressure solver and  implies that error in data propagating through a Poisson solver on a fine enough mesh would not be filtered, and the error propagation is not affected by the fundamental features of the flow domain (i.e., size, shape, dimension, and configuration of the BCs). 
The analysis in the present work (e.g., \eqref{eq: Ar(n) - DD} and \eqref{eq: Ar(n) - DN} for 1D) can be considered a `global' result that takes the fundamental features of the flow domain into account and complements the `local' analysis in \citet{de2012instantaneous}, and is independent of the numerical implementation of the pressure solver. Invoking $\lambda_x\sim 1/n$, $$Ar(\lambda_x) = C\lambda_x^2,$$ where $C$ is a constant depending on the fundamental features of the domain.

\section{Impact of the location of the error}
\label{sec: impact of  Ef location}
In the context of the buckling beam, the location of a concentrated load (bending moment or transverse load) coupled with the configuration of the supporting mechanism determines the deflection of a beam of a given length. Recalling the analogy between the beam deflection~($X$) and the error in the calculated pressure field~($\epsilon_p$), a similar observation is expected for the error propagation for V-Pressure calculations. We specifically wish to observe how the location of the error, coupled with the fundamental features of the domain, can affect the error level in the calculated pressured field.

We can directly perform this test by letting $\epsilon_f$ be a sharp peak while constraining the power of $\epsilon_f$ to be finite. For example, the following rectangular peak can be used: 
\begin{equation}
    \label{eq:rectfun}
     \epsilon_f = \delta_\Pi(x-x_0) =
  \begin{cases}
    \xi,    & |x-x_0| \le \frac{\Pi}{2} \\
    0,                  & |x-x_0| > \frac{\Pi}{2} 
  \end{cases},
\end{equation}
where $\delta_\Pi(x-x_0)$ is a rectangular pulse \textcolor{black}{function}\footnote{\color{black} The particular profile of the peak error may affect the specific value of $Ar(x_0)$. However, this analysis aims to investigate the relative $Ar$ for a given location $(x_0)$ and setup of boundary conditions. Thus, a consistent error peak function is good enough for this `qualitative' analysis.} with a small pulse width $\Pi$ centered at $x=x_0$. The resulting $\epsilon_p$ and the corresponding error amplification ratio ($Ar$) for each $x_0 \in (0,L)$ then can be evaluated by solving \eqref{eq:ErrorPropagation}. $Ar=Ar(x_0)$ is thus a function of the location of the concentrated error.  $Ar(x_0)$ can be considered as a quantification of how sensitive the pressure solver is to the error at a given location. A high value of $Ar(x_0)$ means that pressure field reconstruction is sensitive to the error concentrated  at $x=x_0$. In practice, one would try to reduce the error in the data at  locations with high $Ar(x_0)$ to improve the overall quality of the reconstructed pressure. 

\begin{figure}[h!]
	\centering 
	\includegraphics[width = 0.9\textwidth]{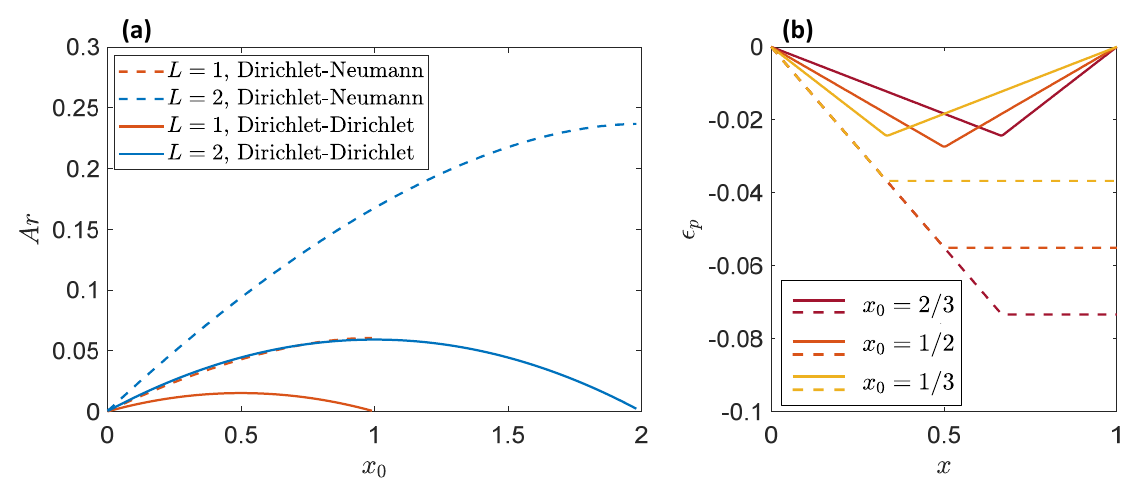}
	\caption{Error amplification ratio ($Ar$) and error profile in the pressure field $\epsilon_p$ when concentrated error ($\epsilon_f = \delta_{0.01L}(x-x_0)$ is introduced at different locations~($x_0$). (a) $Ar$~corresponding to the location of concentrated error~($x_0$) for domains with different sizes and configurations of boundary condition. (b)~Profiles of $\epsilon_p$ over a domain with size $L=1$, when concentrated error $\epsilon_f=\delta_{\Pi}(x-x_0)$ is introduced at $x_0=2/3, 1/2, 1/3$. The solid curves indicate the results on the domain with Dirichlet-Dirichlet BCs, and the dashed curves indicate the results on the domain with Dirichlet-Neumann BCs, respectively.}
	\label{fig:Ar(x0) 1D}
\end{figure}

Figure~\ref{fig:Ar(x0) 1D}(a) demonstrates a case study of the error amplification ratio $Ar(x_0)$ caused by error concentrated at different places in a 1D domain with different sizes. The concentrated error source is constructed using \eqref{eq:rectfun} with width $\Pi=0.01L$ and height  $\xi=10$. The solid curves, representing the Dirichlet domain, peak at the center of each of their domains, and vanish at the Dirichlet boundaries. This indicates that pressure reconstruction is more sensitive to the error located far away from Dirichlet boundaries. This is reflected in Figure~\ref{fig:Ar(x0) 1D}(b) where solution error with the highest peak is associated with the location in the middle of the domain $x_0=0.5$. In the context of a bending beam, a bending moment applied at the center of a simply supported beam deforms the beam the most.

The dashed curves, representing the domain with mixed BCs, in Figure~\ref{fig:Ar(x0) 1D}(a) show that $Ar(x_0)$ is maximized at the Neumann boundary where $x_0=L$, and is significantly higher than the Dirichlet-Dirichlet case. This means that a Neumann BC makes the pressure reconstruction, in general, more sensitive to error in the data and particularly sensitive to any error near the Neumann boundary. This is reflected in Figure~\ref{fig:Ar(x0) 1D}(b) where the $\epsilon_p$ profile grows at a constant slope starting from the Dirichlet boundary until the location of the concentrated error in $\epsilon_f$, after which $\epsilon_p$ plateaus. A concentrated error close to the Neumann boundary leads to a significantly higher $\epsilon_p$. For beam theory, a beam supported by a slide (which is a Neumann 
BC, see Figure~\ref{fig:beam}(b)) at one end is much less resistant to deflection than a simply supported beam, and a moment near the slide end can cause a large deflection of the beam. The error in the pressure field affected by the location of concentrated error in the data can also be analyzed using Green's function in a more rigorous fashion, and $A(x_0)$ can in turn be captured using the  principle of superposition. We will demonstrate this Green's function based analysis with 2D examples in \S\ref{sec:2D examples}. 

Figure~\ref{fig:Ar(x0) 1D} also demonstrates the effect of the size of the domain on the error amplification. 
It is seen that the blue curves (solid or dashed) are higher than the the corresponding red curves, confirming  again that a larger domain tends to amplify the error in the data proportionally to the domain's size. This observation agrees with the conclusion in the previous sections and \citet{Pan2016Error}. In summary, the pressure reconstruction problem is sensitive to the location of the error in the data, depending on the specific fundamental features of the domain.

\section{Case study in 1D}
\label{sec:1D example}
We now carry out a case study in 1D to further demonstrate the results in the above sections.
Consider, for example, a normalized flow profile along the center line of a steady Poiseuille flow ($x \in [0,L] $) driven by a pressure gradient ($dp/dx = -1$), whose velocity profile is defined as $u(x) = 1$. The corresponding V-pressure reconstruction problem is governed by a differential equation (i.e., $d^2p/dx^2 = 0$) by a Poisson approach, which can be solved with proper BCs; for example  Dirichlet-Dirichlet boundaries (i.e., $p(0) = 0$,  and $p(L) = -L$) or Dirichlet-Neumann boundaries (i.e., $p(0) = 0$,  and $p'(L) = -1$). With error absent in the data field (i.e., $\epsilon_f=0$), the pressure profile from solving the Poisson type equation should give $p = -x$. However, any error in the data field ($\epsilon_f$) will propagate to the calculated pressure field and yield nonzero~$\epsilon_p$. 

\begin{figure}[!h]
	\centering 
	\includegraphics[width=\textwidth]{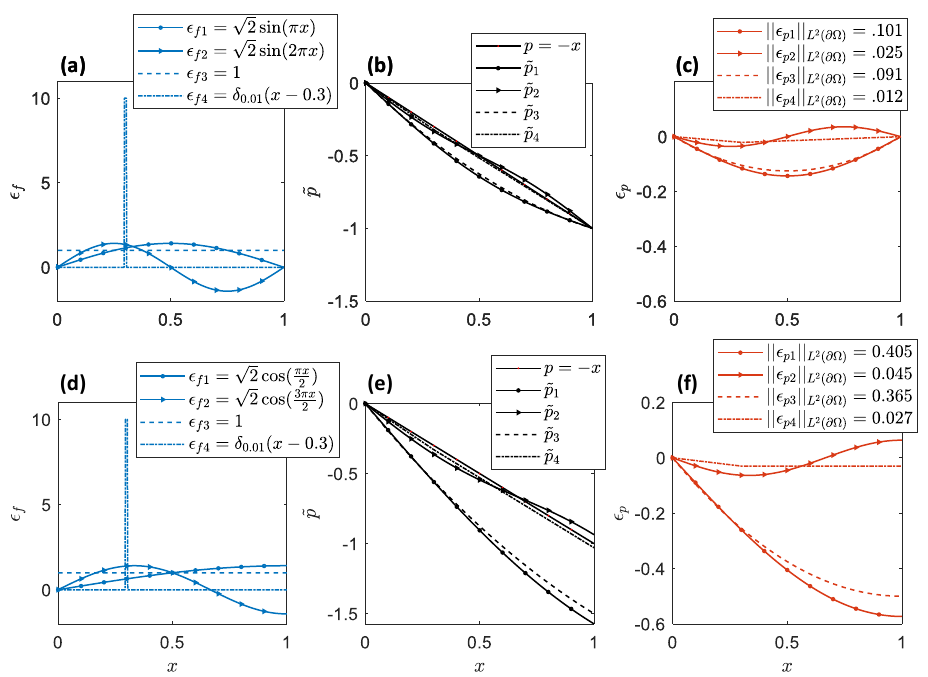}
	\caption{One dimensional case study on the impact of the error profile to the error propagation of the pressure reconstruction problem for a domain with different BCs (upper row (a--c) for Dirichlet-Dirichlet BCs and lower row (d--f) for Dirichlet-Neumann BCs). (a, d)~Four typical profiles of error in the data ($\epsilon_{f}$); expressions are shown in the legend. 
	(b, e)~Reconstructed pressure profile~($ \tilde{p}$) using the data contaminated by the error profiles in (a) and (d), respectively, compared with the exact value ($p=-x$). (c, f)~Error profiles in the calculated pressure ($\epsilon_p$). The error level for each case  ($||\epsilon_p||_{L^2(\Omega)}$) in the calculated pressure field are listed in the legend.}
	\label{fig:1DExample}
\end{figure}

We choose a few typical $\epsilon_f$ to demonstrate how the error profile (particularly, spatial frequency and location of the error), together with the fundamental features of the flow domain, affects the error propagation of the pressure reconstruction (see Figure~\ref{fig:1DExample}, where the pressure reconstruction problems equipped with Dirichlet-Dirichlet and Dirichlet-Neumann BCs are illustrated in the upper and lower row, respectively). 
In Figure~\ref{fig:1DExample}(a) and (d), we illustrate the profiles of the error in the data. $\epsilon_{f_1}$ and $\epsilon_{f_2}$ are the worst and the second worst error in the data for the particular domain, respectively, calculated using \eqref{eq:modes in Ef}. 
$\epsilon_{f_3}$ is a uniform error.  $\epsilon_{f_4}$ is a sharp rectangular impulse centering at $x=0.3$. All these error profiles ($\epsilon_{f_i}$, i = 1,2,3,4) have unit power (i.e., $||\epsilon_{f_i}||_{L^2(\omega)} = 1$). Figure~\ref{fig:1DExample}(b) and (e) show the reconstructed pressure field when the data is contaminated by the error demonstrated in Figure~\ref{fig:1DExample}(a) and (d), respectively, compared with the exact pressure profile ($p=-x$). In Figure~\ref{fig:1DExample}(c) and (f), we show the profiles of the error in the reconstructed pressure field, where the legends list the power of the corresponding error. 

% Figure~\ref{fig:1DExample} illustrates four  different enforced error profiles in the data, where $\epsilon_{f1}$ is the computed maximal error (Fig~\ref{fig:1DExample}(a)) inducing profile (Fig~\ref{fig:1DExample}(c)).
The results shown in Figure~\ref{fig:1DExample} verify the analysis in the previous sections and also provide practical insights. For example, the high frequency error in the data given by $\epsilon_{f2}$ has the same amplitude as $\epsilon_{f1}$ ($\max |\epsilon_{f1}| = \max |\epsilon_{f2}| = \sqrt{2}$), and the power is also the same  ($||\epsilon_{f1}||_{L^2(\Omega)} =  ||\epsilon_{f3}||_{L^2(\Omega)} = 1$) in Figure~\ref{fig:1DExample}(a). However, the induced error in the pressure calculation ($||\epsilon_{p2}||_{L^2(\Omega)}$ = 0.025) is significantly less than the worst case ($||\epsilon_{p1}||_{L^2(\Omega)} = 0.101$, see Figure~\ref{fig:1DExample}(c)). This implies that high frequency errors (e.g., spurious vectors and peak-locking in PIV could be one of the typical sources) affect the error propagation less than the low frequency errors of the same magnitude. $\epsilon_{f4}$ has a sharp tall peak which may be indicative of a local error such as due to spurious vectors. This error is concentrated on a small spatial scale and is smoothed out by the Poisson operator (not necessarily reliant on the numerical scheme as reported by \citet{de2012instantaneous}) so that $\epsilon_{p4}$ is significantly less than the maximal error. In contrast, global low amplitude errors such as $\epsilon_{f3}$ (could be due to erroneous calibration) can yield relatively large error in the pressure calculation ($\epsilon_{p3}$). This example gives a possible reason why low-pass filters of the PIV post-processing do not necessarily always improve PIV-based pressure solvers (reported observation of, for example, \citet{charonko2010assessment}): a low-pass filter may effectively  remove high frequency errors (and signals) in the data, however, the low frequency components of the error which are more influential in terms of error propagation remains and could significantly contaminate the reconstructed pressure field.
The power of the error in the data is constrained to unity ($||\epsilon_{fi}||_{L^2(\Omega)} =1$), hence, the error amplification ratio $Ar$ is numerically equal to the error level in the calculated pressure ($|| \epsilon_{pi} ||_{L^2(\Omega)}$) as shown in the legend of Figure~\ref{fig:1DExample}(c). Note, this example shows that the total error can vary by an order of magnitude even when the errors in the data have the same amplitude (e.g., $\epsilon_{f_1}$ and $\epsilon_{f_2}$); and high amplitude error in the data (e.g., comparing $\epsilon_{f_4}$ and $\epsilon_{f_3}$) does not necessarily lead to high error in the calculated pressure field. The reason is that besides the power or the amplitude of the error in the data, the \textit{profile} of the error plays a large role in the error propagation dynamics of the pressure field calculation. 

These observations also hold for the Dirichlet-Neumann case (see Figure~\ref{fig:1DExample}(d--f)). Comparing \ref{fig:1DExample}(c) and (f), which show the error in the calculated pressure field for Dirichlet-Dirichlet and Dirichlet-Neumann cases, we again note that the presence of the Neumann BC significantly increases the sensitivity to the error in the data field. The maximum error appearing for the Dirichlet-Neumann case is significantly higher compared to the Dirichlet-Dirichlet case. 

In summary, this 1D example confirms that the  profile of the error in the data significantly affects the error in the calculated pressure field. The result concurs with some findings by \citet{charonko2010assessment,Pan2016Error}, and extends the results \citet{de2012instantaneous}, demonstrating that the effects are a result of the Poissson operator, and not necessarily only the numerical scheme used to approximate it.

\section{2D examples}\label{sec:2D examples}
Following the presentation in \S\ref{sec: impact of frequency of Ef}--\ref{sec:1D example}, we provide examples in 2D to demonstrate how the profile of the error (specifically, spatial frequency and location of $\epsilon_f$) in the data field affect the error propagation from $\epsilon_f$ to $\epsilon_p$. A case study in a more practical context is also provided.

\subsection{Effects of spatial frequency of $\epsilon_f$ in 2D}
Following the same procedure demonstrated in \S\ref{sec:Analogy - PIVPressure}, we can analyze 2D domains and calculate the worst error ($\epsilon_f^*$) and the worst amplification ratio ($Ar^*$) in 2D. Consider now a pressure field that is reconstructed on an $N \times M$ rectangular domain, with aspect ratio $\alpha = N/M$ and area $A=MN$, by solving a pressure Poisson's equation with Dirichlet boundaries. Solving the eigenvalue problem \eqref{eq:4thOrderGeneral} in 2D we can find the error amplification ratio
\begin{equation}
\label{eq:2D Ar}
    Ar = \beta^2_{m,n} = \frac{A}{\pi^2(\alpha m^2+\alpha^{-1} n^2)}, \quad  m,n = 1,2,3,...
\end{equation}
where $\beta_{m,n}$ are eigenvalues. The corresponding error in the data field is 
  \begin{equation}
     \label{eq:Ef-DDDD}
     \epsilon_f = 2\sin\left(\frac{m\pi x}{M}\right)\sin\left(\frac{n \pi y}{N}\right).
 \end{equation}
When $m=n=1$, the resultant maximum error amplification ratio is
  \begin{equation}
     \label{eq:Ar*-DDDD}
    Ar^* =  \frac{A\alpha}{\pi^2(1+\alpha^2)}, 
 \end{equation}
which is generated by the worst error in the data that takes the profile:
  \begin{equation}
     \label{eq:Ef*-DDDD}
   \epsilon_f^* =  2\sin\left(\frac{\pi x}{M}\right)\sin\left({\frac{\pi y}{N}}\right).  
 \end{equation}
The maximum amplification ratio in~\eqref{eq:Ar*-DDDD} is identical to the coefficient (Poincare constant) in front of the error level in the data field ($||\epsilon_f||_{L^2(\Omega)}$) for the upper bound of the error in the pressure field (see equation~(13) in \citet{Pan2016Error}, which is re-listed here for convenience):
\begin{equation*}
    ||\epsilon_p||_{L^2(\Omega)}\le \frac{\alpha}{\pi^2(1+\alpha^2)}A ||\epsilon_f||_{L^2(\Omega)} + ||\epsilon_h||_{L^\infty(\partial \Omega)}=Ar^*||\epsilon_f||_{L^2(\Omega)} + ||\epsilon_h||_{L^\infty(\partial \Omega)},
\end{equation*}
where $||\epsilon_h||_{L^\infty(\partial \Omega)}$ measures the error level on the boundary. This means that the upper bound of the error in the calculated pressure is achieved when the error in the data field takes the profile of $\epsilon_f^*$ as described by \eqref{eq:Ef*-DDDD}, for a domain where the contribution from potentially contaminated boundaries can be neglected. The second worst mode of the error in the data field is obtained from $m=1,n=2$ or $m=2,n=1$, and so on.

Similar to the pure Dirichlet BCs case, for an $N \times M$ flow domain with mixed BCs, we can calculate the worst profile of the error in the data field. Consider a pressure field reconstructed, with the same aspect ratio $\alpha$ and area $A$ as the previous example, by solving a pressure Poisson's equation with Dirichlet boundaries on the top and bottom of the domain ($y= 0$ and $y=N$), and Neumann conditions on the left and right boundaries ($x= 0$ and $x=M$). Solving the eigenvalue problem \eqref{eq:4thOrderGeneral} leads to the error amplification ratio taking the same form as \eqref{eq:2D Ar}: $Ar = \beta^2_{m,n} = \frac{A}{\pi^2(\alpha m^2+\alpha^{-1} n^2)}, n = 1,2,3,...$, however, $m = 0,1,2,...$ starts from $0$ due to the Neumann BCs.
% The corresponding error in the data field is
%   \begin{equation}
%      \label{eq:Ef-DDDD}
%      \epsilon_f = C \cdot \cos{\frac{m \pi x}{M}}\sin{\frac{n \pi y}{N}},
%  \end{equation}
% where $C = \sqrt{2}$ for $m = 0$, and $C = 2$ for $m\ge1$. 
The maximum error amplification ratio is
  \begin{equation}
     \label{eq:Ar*-DNDN}
    Ar^* =  \frac{A \alpha}{\pi^2}
 \end{equation}
for $m=0, n=1$, and the corresponding worst error profile in the data is
  \begin{equation}
     \label{eq:Ef*-DNDN}
   \epsilon_f^* =    
    \sqrt{2} \sin\left({\frac{\pi y}{N}}\right).
 \end{equation}
Note again, \eqref{eq:Ar*-DNDN} is identical to the Poincare constant in front of the power of the error in the data field in inequality (19) in \cite{Pan2016Error}. One may note that with the Neumann boundaries present, $Ar^*$ for the domain with mixed BCs (i.e., \eqref{eq:Ar*-DNDN}) is $(1+\alpha^2)$ times greater than $Ar^*$ for the domain with all Dirichlet boundaries (i.e., \eqref{eq:Ar*-DDDD}), implying that Dirichlet boundaries can tame the error propagation, while long Neumann boundaries (corresponding to large $\alpha$ in this setup) can increase the error amplification. In the context of the vibration of elastic bodies, one can think of how a plate simply supported on all the edges is more resistant to deflection than a plate simply supported on only two opposite sides.

\begin{figure}[!h]
	\centering 
	\includegraphics[width=0.65\textwidth]{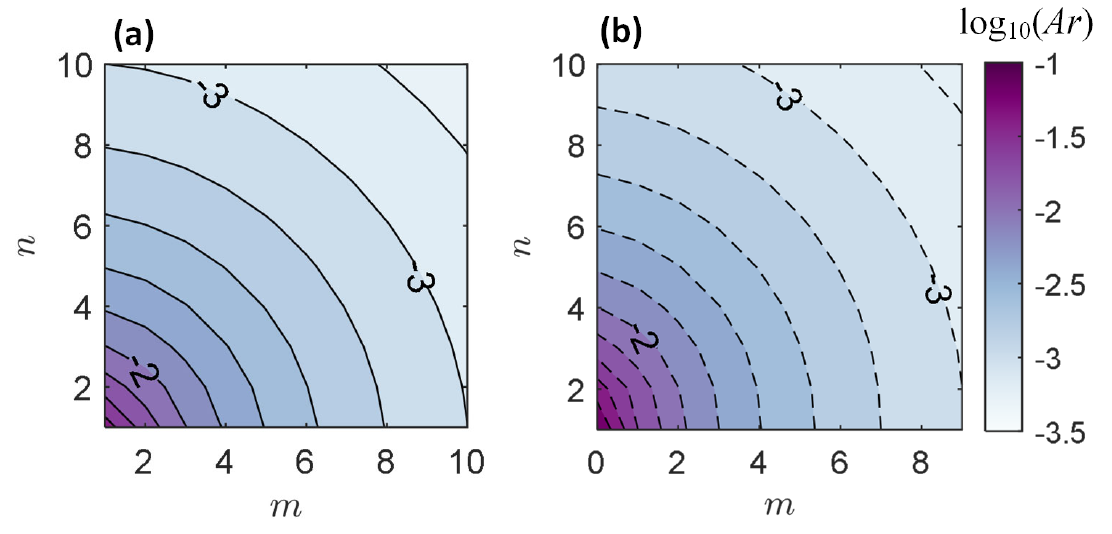}
	\caption{Error amplification ratio responding to spatial frequencies for a domain with pure Dirichlet BCs (a) and mixed BCs (b), respectively.}
	\label{fig:Ar_mn}
\end{figure}

In both the pure Dirichlet and mixed BCs, $Ar$ is inversely proportional to spatial frequency in both the $x$ and $y$ directions \eqref{eq:2D Ar}. Figure~\ref{fig:Ar_mn} demonstrates the error amplification ratio $Ar$ as a function of the spatial frequencies of $\epsilon_f$ for a domain with different BCs. When the spatial frequencies of $\epsilon_f$ are low in either direction, $Ar$ is relatively high. $Ar$ is slightly higher for a given set of frequencies in mixed boundaries than in pure Dirichlet (see \eqref{eq:Ar*-DDDD} and \eqref{eq:Ar*-DNDN}), as shown by Figure~\ref{fig:Ar_mn}(b) being darker than Figure~\ref{fig:Ar_mn}(a) for a given location. High frequency errors are largely filtered for both the Dirichlet case (see  Figure~\ref{fig:Ar_mn}(a)) and the mixed boundary case as well (see Figure~\ref{fig:Ar_mn}(b)).

\subsection{Effects of the location of $\epsilon_f$ in 2D}
\label{sec:Effects of the location of ef in 2D}
We now show the impact of the location of a concentrated error to the calculated pressure field in 2D. Similar to the test in 1D (see \S\ref{sec: impact of  Ef location}), the error sensitivity map can be found by introducing concentrated error source in the data field at different locations: 
\begin{equation}
    \label{eq:rectfun2D}
     \epsilon_f = \delta_\Pi(\bm{x}-\bm{x_0}) =
  \begin{cases}
    \frac{2}{\sqrt{\pi}\Pi},    & |\bm{x}-\bm{x_0}| \le \frac{\Pi}{2} \\
    0,                  & |\bm{x}-\bm{x_0}| > \frac{\Pi}{2} 
  \end{cases},
\end{equation}
where $\bm{x_0}=(x_0,y_0)$ is the coordinate of the center of the concentrated error. The width of the error is chosen to be $\Pi=0.02/\sqrt{|\Omega|}$. As the power of the error is $||\epsilon_f||_{L^2(\Omega)} = 1$, the resulting error in the pressure field is $\epsilon_p$, and its power is numerically equal to~$Ar$.

% Give a short analysis about this Figure~\ref{fig:Ar_location_2D}. Try to recall the 1D case and the 2D examples for Figure~\ref{fig:Ar_mn_DDDD}, and Figure~\ref{fig:Ar_mn_DNDN} and .

\begin{figure}[!h]
	\centering 
	\includegraphics[width=1.0\textwidth]{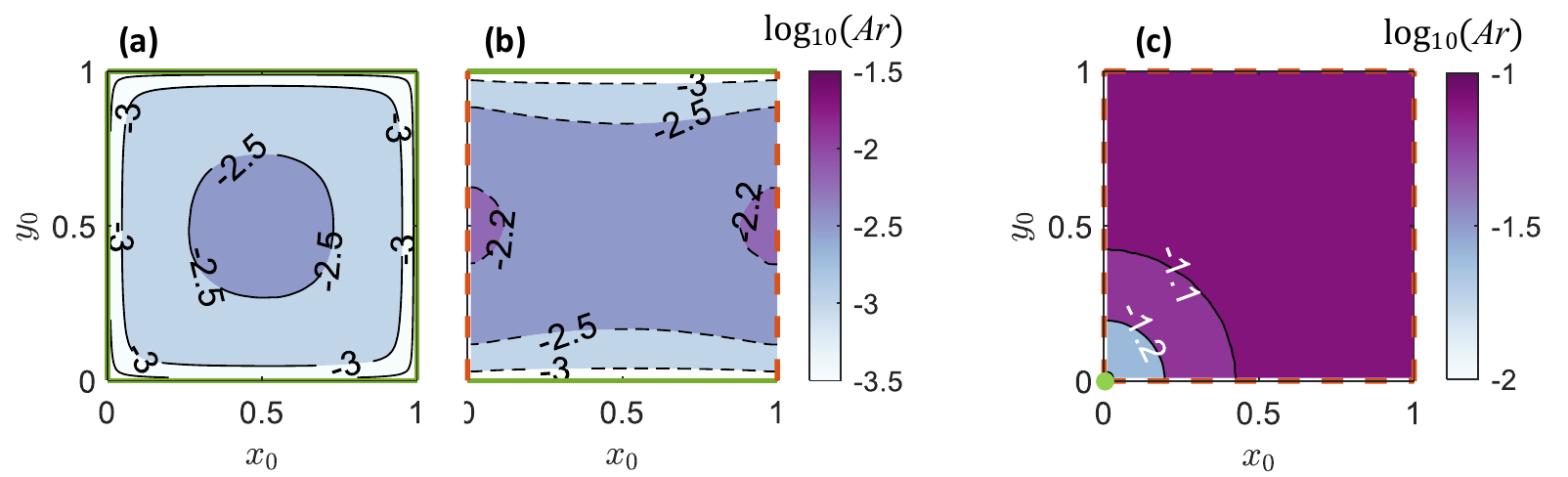}
	\caption{Error amplification ratio $Ar(x_0,y_0)$ responding to concentrated error located at $\bm{x_0}=(x_0,y_0)$ for domains with \textcolor{black}{(a) pure Dirichlet boundaries and (b)\&(c) mixed boundaries, respectively. Dirichlet boundaries are marked by green solid lines in (a)\&(b), and the green dot in (c)}. Neumann boundaries are marked by orange dashed lines.}
	\label{fig:Ar_location_2D}
\end{figure}

Figure~\ref{fig:Ar_location_2D} shows the error sensitivity map on a $1\times 1$ domain for \textcolor{black}{three} distinct BC configurations, where $Ar(x_0,y_0)$ is a function of the location of the concentrated error source as illustrated.
We again show that error near a Dirichlet boundary is less sensitive in terms of error propagation, while the error near a Neumann boundary and/or far away from a Dirichlet boundary is more dangerous. 
In the context of the elastic mechanics of a bending plate, we recall that moments applied next to a simply supported edge do not deflect the plate significantly due to the relatively `firm' support nearby; however, moments exerted near a slide support would cause more deflection over the entire plate. 
% \textcolor{blue}{This is easily observed in Figure~\ref{fig:Ar_location_2D}(c), where the 'sliding' Neumann boundaries allow error to propagate significantly everywhere in the domain, and the singular 'firm' point Dirichlet boundary only slightly reduces the error.}
\textcolor{black}{We further demonstrate this analogy using an `extreme' case shown in Figure~\ref{fig:Ar_location_2D}(c), where a point Dirichlet boundary is applied at $\bm{x_0} =(0,0)$, and all the remaining boundaries are Neumann conditions. In this case, the square plate is held by  `sliding' supports almost everywhere on the edges. This weak configuration leads to a setup that can be significantly buckled by a peak load applied almost everywhere in the domain, except for the region near $\bm{x_0} =(0,0)$, where the deflection is tamed by the `firm' Dirichlet condition. In the context of error propagation of V-pressure reconstruction, a single point Dirichlet BC setup can lead to a problem that is sensitive to error located almost anywhere in the domain, and thus, such a boundary condition setup should be avoided if possible in practice.}

\begin{figure}
    \centering
    \includegraphics[width=0.9\textwidth]{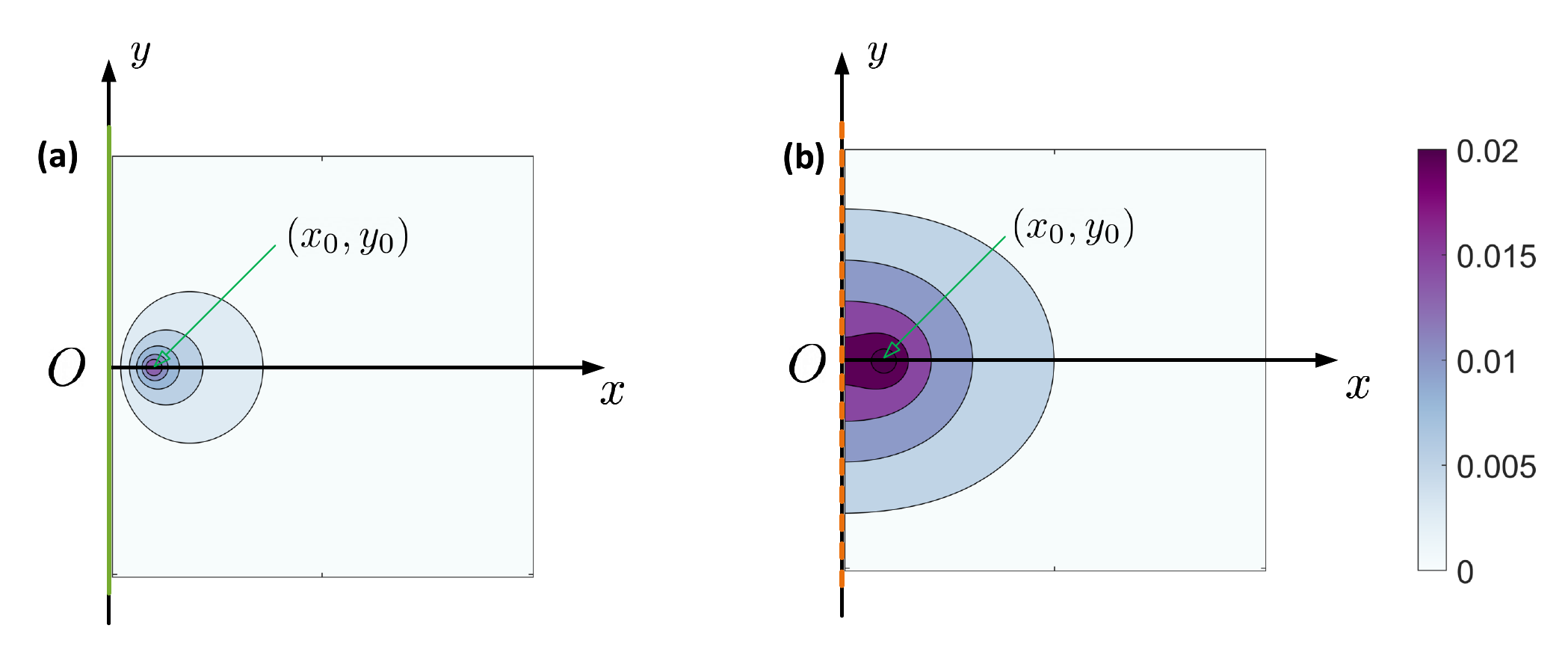}
    \caption{Error in the pressure field ($\epsilon_p$) caused by a concentrated error source in the data field ($\epsilon_f$). $\epsilon_f$ is located at $\bm{x_0}=(x_0,y_0)$ (marked by the green arrow head). The contour map can be considered as an approximation of the Green's function on a half plane $G(\bm{x},\bm{x_0})$, when the location of $\delta(\bm{x},\bm{x_0})$ is close to a homogeneous Dirichlet boundary (a) and a homogeneous Neumann boundary (b), marked by the green solid line and red dashed line at $x=0$, respectively.}
    \label{fig:green func}
\end{figure}

Besides the intuition based on the bending plates, these observations about the effect of the distance from an error source to boundaries with different BCs can be explained using Green's function for the Poisson equation. 
Consider, in 2D, a concentrated error source in the data field taking the form of the Dirac delta function, $\epsilon_f = -\delta(\bm{x_0})$, located at $\bm{x_0}=(x_0,y_0)$ (in the limit as $\Pi\rightarrow 0$ this is the same as the error we have used above). The error in the pressure field caused by this concentrated error source is the fundamental solution of the Poisson equation given by the Green's function. Namely,  $\epsilon_p=G(\bm{x},\bm{x_0})$ such that
\begin{equation}
\nabla^2\epsilon_p = \nabla^2 G(\bm{x},\bm{x_0}) = \delta(\bm{x}-\bm{x_0}) = \epsilon_f. 
     \label{eq: Green's function}
\end{equation}
When the distance from the error to a nearby boundary is much shorter than the distance to any other other boundaries, the local error near the source (and the nearby boundary) can be approximated by the Green's function on a half plane by the method of images. For example, the Green's function on a half plane with a homogeneous Dirichlet boundary at $x=0$ (see Figure~\ref{fig:green func}(a)) is 
\begin{equation*}
\epsilon_p = G(\bm{x},\bm{x_0}) = \frac{1}{2\pi}\ln\left[ \frac{\langle(x,y),(x_0,y_0)\rangle}{\langle(x,y),(-x_0,y_0)\rangle} \right],
    \label{eq:Green's function D}
\end{equation*}
where $\langle(x,y),(x_0,y_0)\rangle = \sqrt{(x-x_0)^2+(y-y_0)^2}$ is the distance from $\bm{x}$ to $\bm{x_0}$ \cite{tikhonov2013equations}. 
As the location of the error approaches the boundary ($x_0 \rightarrow 0$), $G(\bm{x},\bm{x_0})\rightarrow0$ and thus, $\epsilon_p$ vanishes quickly towards the Dirichlet boundary. 

For an error source close to a Neumann boundary (e.g., Figure~\ref{fig:green func}(b)), the corresponding Green's function on the half-plane is 
\begin{equation*}
\epsilon_p = G(\bm{x},\bm{x_0}) = \frac{1}{2\pi}\ln\left[\langle(x,y),(x_0,y_0)\rangle \langle(x,y),(-x_0,y_0)\rangle \right],
    \label{eq:Green's function N}
\end{equation*}
defined up to an undetermined constant, and the maximum error is achieved at the boundary where $(x,y)=(0,y_0)$: 
\begin{equation}
    \label{eq:Green's function N2}
    \epsilon_p  = \frac{1}{2\pi}\ln\left(x_0^2\right),
\end{equation}
which grows fast as $x_0 \rightarrow 0$ (i.e., $\partial\epsilon_p/\partial x_0 \sim x_0^{-1}$). As the location of the concentrated error source approaches the Neumann boundary ($x_0 \rightarrow 0$), $G(\bm{x},\bm{x_0})\rightarrow \infty$ and $\epsilon_p$ blows up at the Neumann boundary. This irregular behavior is not physical, as  $\epsilon_f$ cannot be a strictly Dirac delta function in reality. Despite the  singularity from the analysis using Green's function, the mathematical intuition from this setting shows how the location of the error in the data coupled with the type of BCs affects the error propagation. 

Figure~\ref{fig:green func} shows the contour map of $\epsilon_p$ caused by a concentrated error source  in the data field, where $\epsilon_f=\delta_\Pi(\bm{x} - \bm{x_0})$ is used as an approximation of the delta function. 
% $\bm{x_0}$ is located near a Dirichlet boundary (Figure~\ref{fig:green func}(a)) and a Neumann boundary (Figure~\ref{fig:green func}(b)), respectively. 
It is shown that the same concentrated error source $\epsilon_f$ causes higher amplitude of $\epsilon_p$ in a larger area near a Neumann boundary (Figure~\ref{fig:green func}(b)), compared to the one near a Dirichlet boundary (Figure~\ref{fig:green func}(a)), and consequently leads to a higher error amplification ratio. This is in consensus with the observation in Figure~\ref{fig:Ar_location_2D}. 

\subsection{Case study in 2D}
\begin{figure}[!h]
	\centering 
	\includegraphics[width=0.9\textwidth]{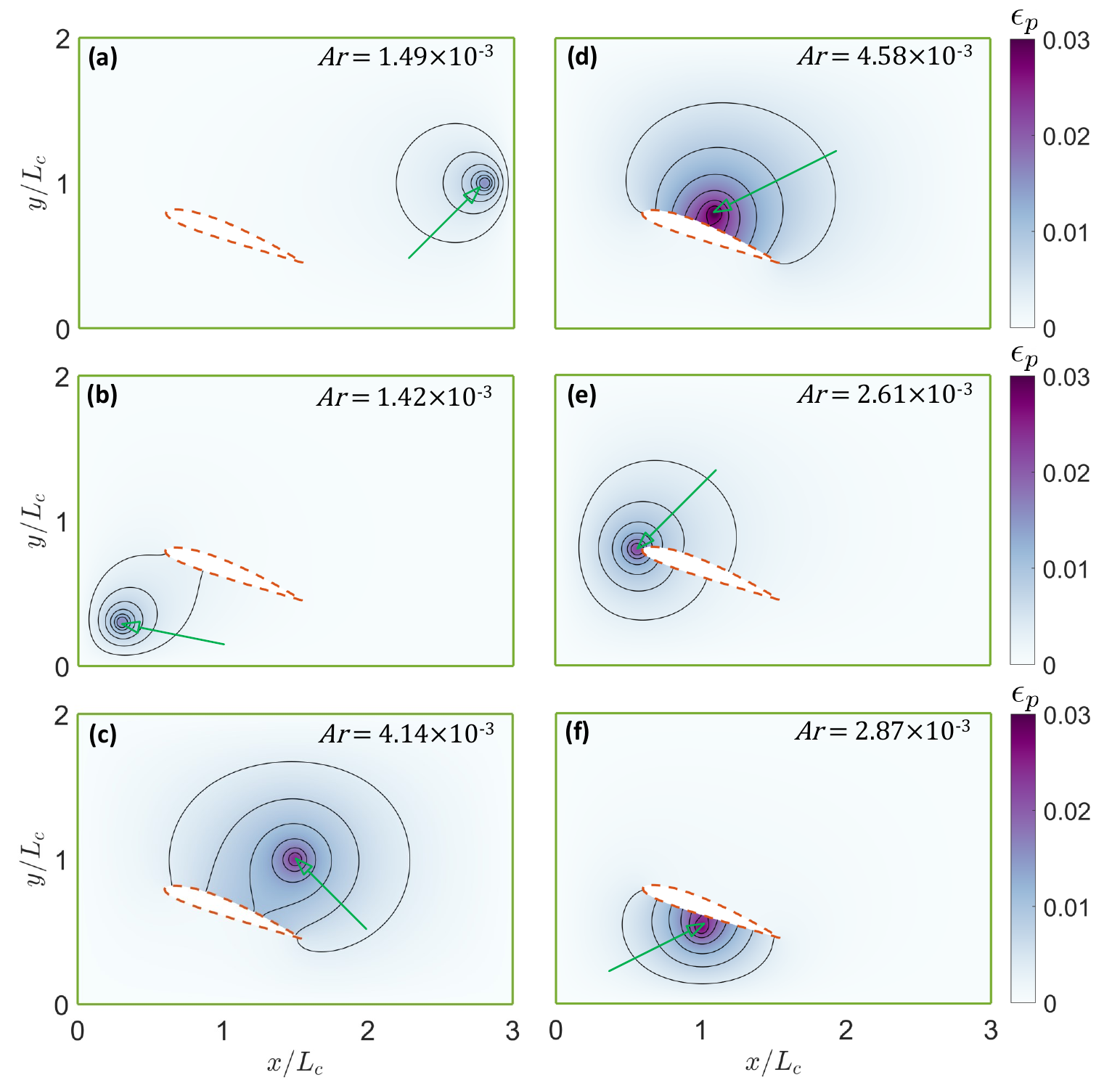}
	\caption{Error in the reconstructed pressure field ($\epsilon_p$) caused by concentrated error in the data field located at different locations ($\epsilon_f=\delta(x_0,y_0)$), for a domain with all Dirichlet BCs on the outer boundaries. The type of BCs of the domain are indicated by green solid lines for Dirichlet boundaries, and dashed red curves for Neumann boundaries. The locations of the error $(x_0,y_0)$ are marked by the green arrow heads.  The legend in each subplot indicates the error amplification ratio ($Ar=||\epsilon_p||_{L^2(\Omega)}/||\epsilon_f||{L^2(\Omega)}$) corresponding to the the $\epsilon_f$ at each location.}
	\label{fig:Ar_mn_DDDD}
\end{figure}

\begin{figure}[!h]
	\centering 
	\includegraphics[width=0.9\textwidth]{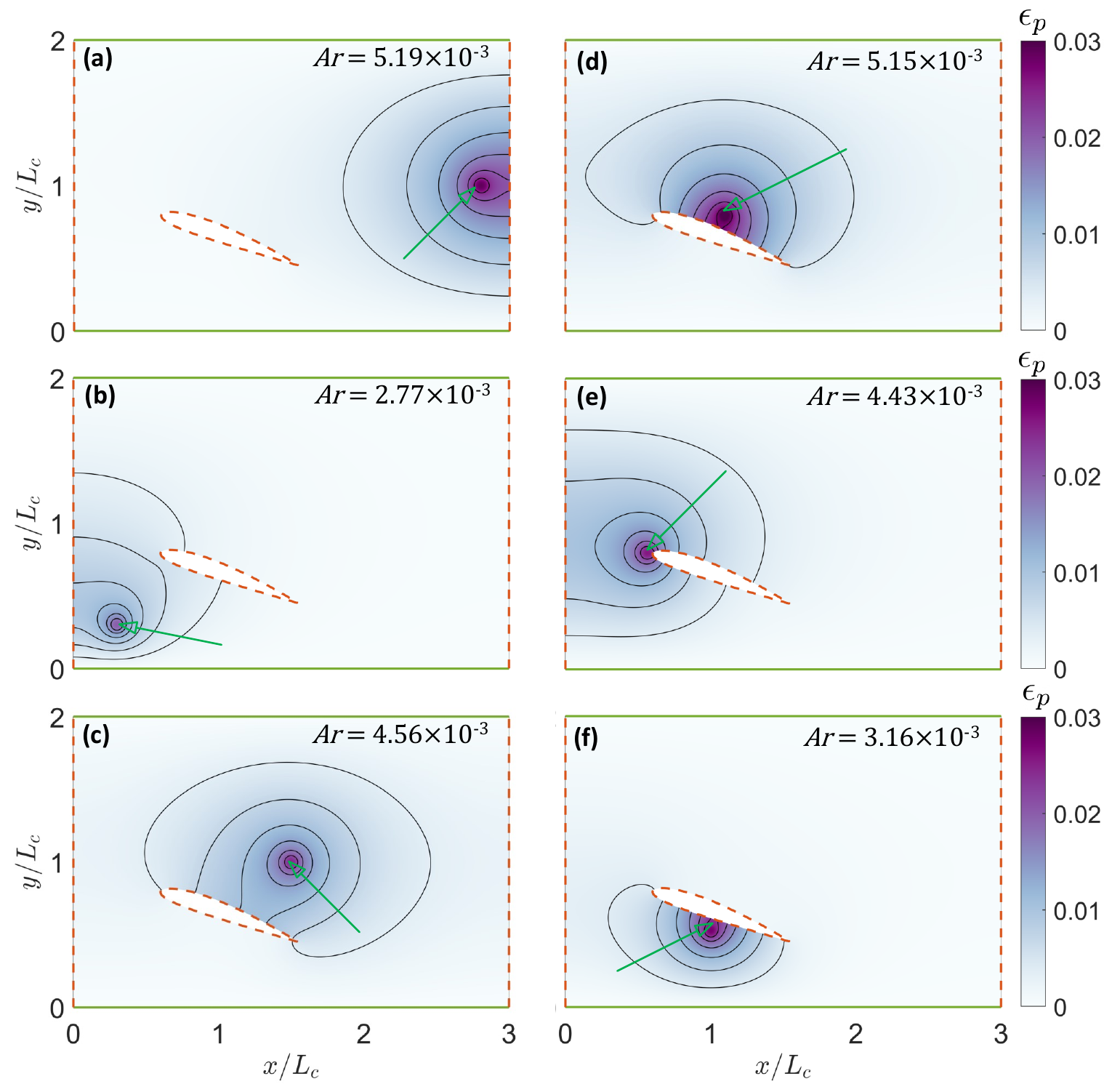}
	\caption{Error in the reconstructed pressure field ($\epsilon_p$) caused by concentrated error in the data field located at different locations ($\epsilon_f=\delta(x_0,y_0)$), for a domain with mixed BCs on the outer boundaries. The type of BCs of the domain are indicated by green solid lines for Dirichlet boundaries, and dashed red lines for Neumann boundaries.  The locations of the error $(x_0,y_0)$ are marked by the green arrow heads. The legend in each subplot indicates the error amplification ratio ($Ar=||\epsilon_p||_{L^2(\Omega)}/||\epsilon_f||{L^2(\Omega)}$) corresponding to the the $\epsilon_f$ at each location.}
	\label{fig:Ar_mn_DNDN}
\end{figure}

To further demonstrate how the location of the error in the data impacts the error propagation, we use a more complex domain, where an airfoil is located in a rectangular domain. The   non-dimensional size, scaled by chord length, of the domain is $N\times M = 3\times2$. Figure~\ref{fig:Ar_mn_DDDD} demonstrates the error propagation in a pure Dirichlet boundary domain and Figure~\ref{fig:Ar_mn_DNDN} is a demonstration for a domain with mixed boundaries.

We first note the effect of the Dirichlet boundaries shown in Figure~\ref{fig:Ar_mn_DDDD}. The error profiles in Figure~\ref{fig:Ar_mn_DDDD}(a) and (b) enjoy the lowest $Ar$ and amplitude of $\epsilon_p$ due to the error in their $\epsilon_f$ functions being concentrated close to Dirichlet boundaries. By contrast, the error profile in Figure~\ref{fig:Ar_mn_DDDD}(c) has error concentrated in the middle of the domain, far from any Dirichlet boundaries, and thus does not experience an error-propagation taming effect from the Dirichlet boundaries, resulting in a higher $Ar$ and high amplitude error in the pressure field ($\epsilon_p$). To use the elastic mechanics analogy of a bending plate, this is equivalent to having difficulty deforming a plate by applying a bending moment at the support, while deforming a plate relatively easily when the bending moment is applied at the center of the plate. 

Neumann boundaries demonstrate the opposite effect, and error in $\epsilon_f$ near Neumann boundaries tends to be more amplified. The case shown in Figure~\ref{fig:Ar_mn_DDDD}(d) has error concentrated closer to the Neumann boundary on the suction surface of the airfoil than in the case for Figure~\ref{fig:Ar_mn_DDDD}(c), and thus has a larger $Ar$, with a larger area significantly contaminated by the error. This can be thought of in terms of a solid mechanics analogy: a moment applied near a sliding support will easily deform a plate. Due to the constraint on the slope of the plate at this boundary, the sliding mechanism will `pull' on other sections of the plate and bring them along with the loaded section. Long Neumann boundaries are more `dangerous' than short ones, as this effect is carried out over a longer distance. The error profile in Figure~\ref{fig:Ar_mn_DDDD}(e) has a concentrated error located next to the leading edge of the airfoil (equivalent to a short Neumann boundary due to its high curvature), and thus has a lower $Ar$ than Figure~\ref{fig:Ar_mn_DDDD}(d), which has a concentrated error near a longer Neumann boundary. It is also worth noting that the error profiles in both Figure~\ref{fig:Ar_mn_DDDD}(e) and (f) benefit from having their concentrated errors relatively close to Dirichlet boundaries, which is why the case shown in Figure~\ref{fig:Ar_mn_DDDD}(f) has less error propagation than the case shown in Figure~\ref{fig:Ar_mn_DDDD}(d) despite both concentrated errors being near Neumann boundaries of similar lengths.
When error in $\epsilon_f$ is concentrated near the suction or pressure surface of an airfoil (see Figure~\ref{fig:Ar_mn_DDDD}(d) and (f), respectively), high amplitude error in the pressure field ($\epsilon_p$) is observed, and the pressure field on the airfoil surface is significantly contaminated. This is especially unfavorable to surface load reconstruction based on velocimetry data, given that load estimation is particularly valuable in engineering applications. 
One effective solution to overcome this uncooperative effect is to add a firm `support' (i.e., accurate Dirichlet BCs) next to the Neumann boundaries. For example, even a few isolated pressure transducers, if placed at the proper locations, can significantly improve the pressure field reconstruction results (\citet{shanmughan2020optimal}).

The coupled effects of BC configurations and the location of concentrated error are further demonstrated in Figure~\ref{fig:Ar_mn_DNDN}. 
The highest $Ar$ values are observed in Figure~\ref{fig:Ar_mn_DNDN}(a) and (d), due to the concentrated errors in these cases being located next to long Neumann boundaries; while when concentrated error is next to a short Neumann boundary as in Figure~\ref{fig:Ar_mn_DNDN}(e), a lower $Ar$ is observed, as expected. The lowest $Ar$ among the tested cases appears in Figure~\ref{fig:Ar_mn_DNDN}(b), where the concentrated error in $\epsilon_f$ is located near a Dirichlet boundary at the bottom of the domain. Note also that the $Ar$ of Figure~\ref{fig:Ar_mn_DNDN}(b) is higher than for Figure~\ref{fig:Ar_mn_DDDD}(b), since the Neumann boundary to the left of the error peak in the mixed case spreads the error further than the Dirichlet boundary in the pure Dirichlet case. Because of the Neumann boundaries on the left and right of the rectangular domain, $Ar$ for the case presented in Figure~\ref{fig:Ar_mn_DNDN}(c) is slightly higher than the one in Figure~\ref{fig:Ar_mn_DDDD}(c); both are relatively high, because their error peaks are far away from any Dirichlet boundaries. Similarly, the $Ar$ for the case shown in Figure~\ref{fig:Ar_mn_DNDN}(e) is increased relative to the pure Dirichlet case (see Figure~\ref{fig:Ar_mn_DDDD}(e)), as the Dirichlet boundary to the left of the error peak is replaced with a Neumann boundary and the taming effect of the Dirichlet boundary is replaced with the more `dangerous' Neumann boundary.
Finally,  $Ar$ for the case shown in Figure~\ref{fig:Ar_mn_DNDN}(f) is slightly higher than the $Ar$ shown in Figure~\ref{fig:Ar_mn_DDDD}(f) due to the presence of Neumann boundaries. However, the $Ar$ increase  is not as large as the cases discussed previously, as the main error taming effect is due to the bottom Dirichlet boundary, which is present in both cases.

\begin{figure}[!h]
	\centering 
	\includegraphics[width=0.9\textwidth]{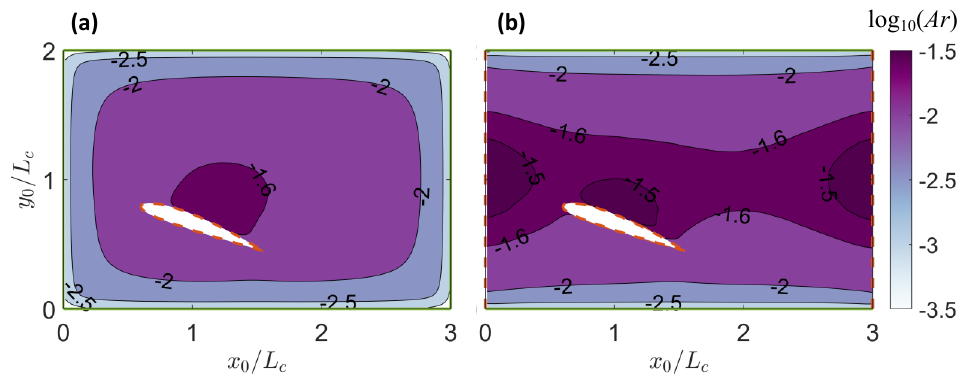}
	\caption{Error amplification ratio $Ar(\bm{x_0})$ responding to concentrated error located at $\bm{x_0}=(x_0,y_0)$ for domains featuring an airfoil of chord length $L_c$. Neumann BCs are applied on the surface of the airfoil. The domains feature (a)  Dirichlet BCs on the outer boundaries and (b) mixed BCs on the outer boundaries. Dirichlet boundaries are marked by green solid lines and Neumann boundaries are marked by orange dashed lines, respectively.}
	\label{fig:Ar_vs_location_airfoil}
\end{figure}

These observations in Figures~\ref{fig:Ar_mn_DDDD} and \ref{fig:Ar_mn_DNDN} are demonstrated further in Figure~\ref{fig:Ar_vs_location_airfoil}, which shows the error sensitivity map for a 3 $\times$ 2 domain with an airfoil, nondimensionalized by the airfoil chord length, for different BC configurations. $Ar(x_0,y_0)$ is the error amplification ratio as a function of the location of concentrated error as illustrated. Figure~\ref{fig:Ar_vs_location_airfoil}(a) shows that concentrated error near a Dirichlet BC is tamed by the boundary, as seen by the low value of amplification ratio $Ar$ around the edges of the domain. Concentrated errors far from Dirichlet boundaries and/or close to a Neumann boundary show more propagation, and thus a higher value of $Ar$ occurs near the center of the domain. Similarly, as show in Figure~\ref{fig:Ar_vs_location_airfoil}(b), a high value of $Ar$ is located around $y_0/L_c =1 $.  In Figure~\ref{fig:Ar_vs_location_airfoil}(b), we see that error amplification is highest in the middle of long Neumann boundaries such as the left and right edges of the domain, and the suction surface of the airfoil. The pressure surface of the airfoil does not suffer from high $Ar$ as it is close to a Dirichlet BC on the bottom of the domain (see both Figure~\ref{fig:Ar_vs_location_airfoil}(a) and (b)).  
% Short Neumann boundaries, such as the front of the airfoil, have less error amplification than long Neumann boundaries. 

% \begin{figure}[!h]
% 	\centering 
% 	\includegraphics[width=0.9\textwidth]{Ar_vs_location_airfoil.pdf}
% 	\caption{Error amplification ratio $Ar(\bm{x_0})$ responding to concentrated error located at $\bm{x_0}$ for domains featuring an airfoil of chord length $L_c$ with Neumann BCs. The domains feature  (a) pure Dirichlet edge boundaries and (b)  mixed edge boundaries, respectively. Dirichlet boundaries are marked by green solid lines and Neumann boundaries are marked by red dashed lines.}
% 	\label{fig:Ar_vs_location_airfoil}
% \end{figure}

\section{Conclusions}
\label{sec:Conclusions}
In the present paper, we seek the error in the data field with a particular profile (called the `worst error'), which is amplified most through V-pressure reconstruction. This problem can be modeled using a variational technique, and the worst error can be found by solving the corresponding Euler-Lagrange equations. We find that the Euler-Lagrange equations raised from the error propagation of V-pressure reconstruction is identical to the eigenvalue problem from the buckling of elastic bodies. Particularly, the worst error in the data for a V-Pressure problem is equivalent to the profile of the bending moment associated with the principle mode of, for example, a Kirchhoff–Love plate in 2D.

Inspired by the theory of vibrating elastic bodies, we further analyze how the profile of the error in the data coupled with the fundamental features of the flow domain fundamentally affects the error propagation in the V-Pressure problem, and potentially the performance of a V-Pressure solver. We show that the low frequency errors in the data field are more amplified through a V-Pressure calculation on a large domain with Neumann boundaries; while high frequency components in the error tend to be filtered out. We emphasize that this low-pass filter effect is inherently raised from the property of the Laplace operator and is subject to the particular setup of the domain (e.g., configuration of boundary conditions, shape, and size of the domains). This observation can be coherently explained by analogy with a bending elastic body. For example, low frequency bending moments (equivalent to low frequency error in the data) exerted on a large plate (equivalent to a large domain) with slide supports on the edges (equivalent to Neumann BCs) typically lead to large deflection over the plate.

We also investigate how a concentrated error source in the data field propagates to the pressure field by introducing error taking the profile of a sharp peak in the data. We first define the ratio of the error level in the pressure field to the error level in the data field at each location as the error amplification ratio, which is a measure of the error sensitivity to the location of the error. 
By interrogating the error amplification ratio at each location in a domain, we obtain an error sensitivity map of the domain. We find that the concentrated error located near a accurate   Dirichlet boundary is mild, while V-Pressure reconstruction is much more sensitive to the error near a Neumann boundary and/or far away from a Dirichlet boundary. The calculated pressure field near a Neumann boundary can be significantly contaminated by minor errors in the measured data. This observation is supported by an analysis based on Green's function of Poisson's equation, in 2D as an example. These observations and analysis are also in consensus with the the behavior of buckling elastic bodies. For example, in 1D, a concentrated load applied near a simply supported end (equivalent to a Dirichlet BC) of a beam causes less deflection on the beam than the same load exerted near an end supported by a slide. The same analysis and analogy apply for higher dimensional problems.

The present paper demonstrates the surprising analogy between the error propagation in V-Pressure calculations and solid mechanics of buckling elastic bodies. This analogy is particularly helpful as it allows experimentalists to qualitatively analyze the error or uncertainty propagation dynamics for a complex problem. For example, the error propagation behavior of the V-Pressure problem on a domain with complex fundamental features contaminated by error in the data with complex profiles can be heuristically analyzed by thinking of the behavior of bending a plate with the corresponding set up. Basic understanding of beam or plate theory and textbooks such as \cite{timoshenko1937vibration} can greatly improve understanding of error propagation and the `worst error' in V-pressure calculations, while also reducing the effort needed to apply the theories in this research to experimental fluid studies.

The current  work is presented in order to demonstrate the complicated behavior of the error propagation dynamics of the V-Pressure problem on different aspects: from heuristic analogy, rigorous analysis, to simple case studies based on simulations. The results can assist experimentalists in designing and optimizing tests that are less sensitive to error in the data. These results also are useful in determining which errors in the data can impact the pressure calculation the most, which can be used to benchmark V-Pressure solvers or algorithms.

\textcolor{black}{Finally, we want to emphasize that the present work focuses on how the error in the data field propagates to the reconstructed pressure field. How to interpret error in the data field ($\epsilon_f$) from the error in the velocity field ($\epsilon_{\bm{u}}$) can be challenging. The error propagation analysis from velocity to data ($\epsilon_{\bm{u}} \rightarrow \epsilon_f$) involves the temporal and spatial resolution of velocimetry and specific numerical schemes that evaluate gradients, and thus is out of the scope of the current analytical research. We will leave this topic for future studies, however, some relevant results can be found in, for example,  \cite{mcclure2019generalized,pan2018error}.}

\section*{Acknowledgements}
We would like to thank Drs. Barton Smith and Tadd Truscott for practical discussions. This work is partially supported by Natural Science and Engineering Research Council (NSERC).

\bibliographystyle{dcu}
\bibliography{ref}
\end{document}